\input epsf.tex    

\documentstyle[12pt]{article}

\newcommand{\bmat}{\left(\begin{array}}
\newcommand{\emat}{\end{array}\right)}

\def\yzero{\smash{\hbox{$y\kern-4pt\raise1pt\hbox{${}^\circ$}$}}}

\def\beq{\begin{equation}}
\def\eeq{\end{equation}}
\def\beqa{\begin{eqnarray}}
\def\eeqa{\end{eqnarray}}

\def\-{\hphantom{-}}
\def\ov{\overline}
\def\s2{\frac{1}{2}}

\def\beq{\begin{equation}}
\def\eeq{\end{equation}}
\def\beqa{\begin{eqnarray}}
\def\eeqa{\end{eqnarray}}
\def\tr{{\rm tr \,}}
\def\Tr{{\rm Tr \,}}

\def\diag{{\rm diag \,}}
\def\IF{\relax{\rm I\kern-.18em F}}
\def\II{\relax{\rm I\kern-.18em I}}

\def\cp{{\cal P}}
\def\IC{\bf C}
\def\IZ{\bf Z}
\def\IR{\bf R}
\def\IY{\bf Y}
\def\IS{\bf S}
\def\IP{\bf P}
\def\IT{\bf T}

\def\IX{\bf X}
\def\z2z2{$\IC^3/(\IZ_2\times\IZ_2)$}

\def\id{{\bf 1}}

\def\NN{{\cal N}}
\def\Dsl{\,\raise.15ex\hbox{/}\mkern-13.5mu D} 

 \def\cp#1{\relax\ifmmode {\IP\kern-2pt{}_{#1}}\else $\IP\kern-2pt{}_{#1}$\=fi}
\newcommand{\drawsquare}[2]{\hbox{%
\rule{#2pt}{#1pt}\hskip-#2pt
\rule{#1pt}{#2pt}\hskip-#1pt
\rule[#1pt]{#1pt}{#2pt}}\rule[#1pt]{#2pt}{#2pt}\hskip-#2pt
\rule{#2pt}{#1pt}}

\newcommand{\fund}{\raisebox{-.5pt}{\drawsquare{6.5}{0.4}}}
\newcommand{\Ysymm}{\raisebox{-.5pt}{\drawsquare{6.5}{0.4}}\hskip-0.4pt%
        \raisebox{-.5pt}{\drawsquare{6.5}{0.4}}}
\newcommand{\Yasymm}{\raisebox{-3.5pt}{\drawsquare{6.5}{0.4}}\hskip-6.9pt%
        \raisebox{3pt}{\drawsquare{6.5}{0.4}}}
\newcommand{\antifund}{\overline{\fund}}

\begin{document}

\pagestyle{empty}
\vspace*{.5in}
\rightline{FTUAM-03/03, IFT-UAM/CSIC-03-07}
\rightline{\tt hep-th/0303024}
\vspace{1.5cm}
 
\begin{center}
\LARGE{\bf 
Chiral 4d string vacua with D-branes and NSNS 
and RR fluxes\\[10mm]}

\medskip

\large{Juan F. G. Cascales$^1$, Angel M. Uranga$^2$} \\
{\normalsize {\em $^1$ Departamento de F\'{\i}sica Te\'orica C-XI \\
and Instituto de F\'{\i}sica Te\'orica, C-XVI \\
Universidad Aut\'onoma de Madrid \\
Cantoblanco, 28049 Madrid, Spain \\
{\tt juan.garcia@uam.es}\\
$^2$ IMAFF and \\
 Instituto de F\'{\i}sica Te\'orica, C-XVI \\
Universidad Aut\'onoma de Madrid \\
Cantoblanco, 28049 Madrid, Spain \\
{\tt angel.uranga@uam.es} \\[2mm]}}

\end{center}
 
\smallskip

\begin{center}
\begin{minipage}[h]{14.5cm}
{\small
We discuss type IIB orientifolds with D-branes, and NSNS and RR field 
strength fluxes. The D-brane sectors lead to open string spectra with 
non-abelian gauge symmetry and charged chiral fermions. The closed string 
field strengths generate a scalar potential stabilizing most moduli. We 
describe the construction of models with $\NN=1$ supersymmetric 
subsectors in the context of orientifolds of IIB theory on 
$\IT^6/(\IZ_2\times \IZ_2)$, containing 
D9-branes with world-volume magnetic fluxes, and illustrate model 
building possibilities with several explicit examples. We comment on 
a T-dual picture with D8-branes on non-Calabi-Yau half-flat geometries,
and discuss some of the topological properties of such configurations.
We also explore the construction of models with fluxes and with D3-branes 
at singularities and present a non-supersymmetric 3-family $SU(5)$ model.
}
\end{minipage}
\end{center}

\newpage                                                        

\setcounter{page}{1} \pagestyle{plain}
\renewcommand{\thefootnote}{\arabic{footnote}}
\setcounter{footnote}{0}

\section{Introduction}

One of the most remarkable features of string theory is that, despite its
complexity, it admits vacua with low-enery physics surprisingly close to
the structure of observed particles and interactions. In particular there
exist by now several classes of setups (e.g. heterotic compactifications,
type II models with D-branes at singularities, or intersecting D-branes,
compactifications of Horava-Witten theory, etc) leading to four-dimensional
gravitational and non-abelian gauge interactions, with charged chiral
fermions. Within each class, particular explicit models with spectrum
very close to that of the (Minimal Supersymmetric) Standard Model have
been constructed.
On the other hand, a generic feature of all these constructions,
is the existence of a (very often large) number of moduli, which remain
massless in the construction, unless some supersymmetry breaking mechanism
is proposed. From this viewpoint these models are relatively far from
describing physics similar to the observed world.

Recently, it has been shown that, in the setup of Calabi-Yau
compactification of type II string theory (or also M-theory), there exists a
natural mechanism which stabilizes most moduli of the compactification.
This is achieved by considering compactifications with non-trivial
field strength fluxes for closed string $p$-form fields. This proposal
has already been explored in different setups, leading to large classes
of models with very few unstabilized moduli. Hence this mechanism is one of
the most interesting recent insights to address the long-standing problems
of moduli in phenomenological string models.
Unfortunately, compactifications
with NSNS and RR field strength fluxes have centered on simple models, which
lead to uninteresting gauge sectors, from the phenomenological viewpoint. In
particular, the class of models studied are naturally non-chiral, since
the corresponding gauge sectors arise from too simple stacks of parallel
D-branes.

Our purpose in the present paper is to construct models which combine
the interesting features from the above two classes of theories. In fact,
we succeed in constructing the first string compactifications with
interesting 4d chiral gauge sectors and flux-stabilization of (most) moduli.
The models are based on introducing NSNS and RR 3-form fluxes in
compactifications of type IIB theory with D-branes. Chirality on the latter
arises from the non-trivial gauge bundles on the D-brane world-volumes
(a mechanism related, in the absence of fluxes, to D-brane intersections
via T-duality). For concreteness, and also to 
simplify the discussion of the stability of the configurations,
we choose to center on models with subsectors with $\NN=1$ supersymmetry 
in 4d. In the setup of D-brane with world-volume magnetic fields, a 
simple suitable background geometry is (an orientifold of)
 the $\IT^6/(\IZ_2\times \IZ_2)$ orbifold. However,
we expect that the techniques and our new observations are useful in
constructing supersymmetric models in other geometries; and also to
the construction of non-supersymmetric models with these features.
In fact, we provide an explicit construction of a 
non-supersymmetric compactification with fluxes with a gauge sector of 
D3-branes at singularities leading to a 3-family $SU(5)$ GUT.

\medskip

The paper is organized as follows. In Section \ref{fluxrev} we review the 
construction of toroidal orientifold compactifications with NSNS and RR 
3-form fluxes, and discuss the main points to be addressed in the 
construction of orbifold models. In section \ref{philosophy} we discuss
diverse classes of D-brane configurations leading to chiral gauge sectors, 
and issues arising in the possible introduction of 3-form fluxes for them.
We conclude that a simple and flexible setup for model building is 
that of D-branes with world-volume magnetic fields. In 
section \ref{magnetised} we review the properties and spectra in 
configurations of such magnetised D-branes in toroidal orientifold models, 
and describe the new features when $\IZ_2\times \IZ_2$ orbifold 
projections are included. 

In Section \ref{themodels} we construct explicit models (with 
supersymmetric subsectors) of compacifications with 3-form fluxes and 
configurations of magnetised D-branes. These models have flux-stabilized 
moduli and include gauge sectors with realistic features, namely 
reasonable gauge groups, and several families of charged  chiral fermions.
These constructions provide the first examples in a presumably rich and 
intereseting class of realistic models. 

In Section \ref{bfield} we describe the main modifications 
in models with a discrete NSNS B-field on some two-torus. We show that
it implies the appearance of positively charged O3- and O7-planes in the 
configuration, hence reducing the RR tadpole, and potentially modifying 
the quantization conditions for 3-form fluxes. In section \ref{noncy} we 
discuss the T-duality relation of our models with compactifications with 
D-branes on non-Calabi-Yau geometries, and obtain some of the topological 
couplings relevant to the construction of the later models. Section 
\ref{final} contains our final comments. 

Appendix \ref{zthree} describes an alternative approach to 
compactifications with chiral gauge sectors and few moduli. We construct
explicit (non-supersymmetric) compactifications containing 3-form fluxes 
and chiral gauge sectors arising from D3-branes at singularities, in an 
orientifold of the $\IT^6/\IZ_3$ orbifold. Finally appendix \ref{ktheory} 
addresses several subtle issues on the properties of D-branes in the
presence of NSNS 3-form fluxes (related to the modifications of the 
K-theory group classifying D-brane states in the presence of 3-form flux).
They involve the need to include additional D-branes in some of the models 
of the main text (not modifying its chiral spectrum), and the possibility 
of D-brane instanton mediated transitions between D-branes and fluxes.

After this paper was finished and was being prepared for submission, we 
noticed \cite{recent}, which also studies the same class of models.

\subsection{The $\IT^6/(\IZ_2\times\IZ_2)$ orbifold}
\label{geometry}

The explicit models we will discuss are based on (an orientifold of) the 
$\IT^6/(\IZ_2\times \IZ_2)$ orbifold with discrete torsion (corresponding 
to Hodge numbers $(h_{1,1},h_{2,1})=(3,51)$).
Let us describe the geometry of the $\IT^6/(\IZ_2\times \IZ_2)$ orbifold
(see \cite{vw}). The $\IZ_2\times \IZ_2$ generators $\theta$, $\omega$
act by
\beqa
\theta & : & (z_1,z_2,z_3)\to (-z_1,-z_2,z_3) \nonumber \\
\omega & : & (z_1,z_2,z_3)\to (z_1,-z_2,-z_3) 
\eeqa
on the complex coordinates of $\IT^6=(\IT^2)^3$.
Namely $z_i\to e^{2\pi iv_i} z_i$ with $v_\theta=(1/2,-1/2,0)$ and
$v_\omega=(0,1/2,-1/2)$.

The spectrum of type IIB string theory on the $\IZ_2\times \IZ_2$ orbifold
contains in the untwisted sector the 4d $\NN=2$ supergravity multiplet,
the dilaton hypermultiplet, three vector multiplets and three
hypermultiplets. The moduli in these vector and hypermultiplets correspond
to the complex structure and Kahler parameters of the three underlying
$\IT^2$'s. Components of the metric $G_{i\ov j}$ mixing different two-tori
are projected out by the orbifold actions. The contribution of untwisted
modes to the Hodge numbers of the orbifold is $(h_{1,1},h_{2,1})_{\rm unt}
=(3,3)$.

The orbifold contains twisted sectors of $\theta$, $\omega$ and 
$\theta\omega$. The fixed point set of each twist is given by 16 two-tori, 
near each of which the Calabi-Yau has a local geometry $\IC^2/\IZ_2\times 
\IT^2$. Geometrically, for our choice of discrete torsion, each of these 
fixed tori leads to two collapsed  3-cycle,  given by the $\IP_1$ 
collapsed at the $\IC^2/\IZ_2$ singularity times the 1-cycles in $\IT^2$.
The 3-cycles are thus of $(2,1)$ and $(1,2)$ type. The contribution of
twisted sectors to the Hodge numbers of the orbifold is
$(h_{1,1},h_{2,1})_{\rm tw.}=(0,3\times 16)=(0,48)$. The Hodge numbers of
the orbifold space are therefore $(h_{1,1},h_{2,1})=(3,51)$.

\section{Review of fluxes}
\label{fluxrev}

Compactifications of type II theories (or orientifolds thereof) with
NSNS and RR field strength fluxes have been considered, among others, in
\cite{fluxes,gkp,fp,kst,tt,ferrara}. In this section we review properties 
of type IIB compactifications with 3-form fluxes.

\subsection{Consistency conditions and moduli stabilization}

Type IIB compactification on a Calabi-Yau threefold $X_6$ with
non-trivial NSNS and RR 3-form field strength backgrounds $H_3$, $F_3$
have been extensively studied. In particular the analysis in \cite{gkp} 
provided in a very general setup the consistency conditions such fluxes 
should satisfy. They must obey the Bianchi identities
\beqa
dF_3=0 \quad dH_3=0
\eeqa
and they should be properly quantized, namely for any 3-cycle
$\Sigma\subset {\bf X}_6$
\beqa
\frac{1}{(2\pi)^2\alpha'} \, \int_{\Sigma} F_3 \, \in {\IZ} \quad ; 
\quad
\frac{1}{(2\pi)^2\alpha'} \, \int_{\Sigma} H_3 \, \in {\IZ}
\eeqa
The fluxes hence define integer 3-cohomology classes in $H^3(X_6,\IZ)$.

A subtlety in flux quantization in toroidal orientifolds was noticed in 
\cite{fp,kst}, see also section \ref{bfield}. 
Namely, if flux integrals along some 3-cycle are integer 
but odd, consistency requires the corresponding 3-cycle to pass through an 
odd number of exotic O3-planes (studied in \cite{witbar}). For simplicity 
we restrict to the case where all flux integrals are even integers.

An important observation in \cite{gkp} is that, in order to avoid previous 
no-go theorems about the existence of configurations of fluxes satisfying 
the equations of motion, it is crucial to include orientifold 3-planes in 
the compactification, so we consider type IIB orientifolds with these 
objects. The simplest way to understand the need of these objects, is to 
notice the type IIB supergravity Chern-Simons coupling
\beqa
\int_{M_4\times X_6}\, H_3 \wedge F_3 \wedge C_4
\label{cs}
\eeqa
where $C_4$ is the IIB self-dual 4-form gauge potential. This coupling 
implies that upon compactification the flux background contributes to a 
tadpole for $C_4$, with positive coefficient $N_{\rm flux}$ 
\footnote{This is so if we require the flux to preserve the same 
supersymmetry as the O3-planes. This is implicit in the literature, since 
fluxes leading to negative RR 4-form charge would lead to uncancelled NSNS 
tadpoles, and hence to non-Poincare invariant 4d theories.} (in D3-brane 
charge units). Moreover, fluxes contribute positively to the energy of the 
configuration, due to the 2-form kinetic terms. 
The only way to cancel these tadpoles is to introduce objects with 
negative RR $C_4$-charge and negative tension, to cancel both the RR 
tadpole and also to compensate the vacuum energy of the configuration. 
Having O3-planes in the configuration, it is natural to consider the 
possibility of adding $N_{Q_3}$ explicit D3-branes as well. The RR tadpole 
cancellation constraint hence reads
\beqa
N_{Q_3}\, + \, N_{\rm flux}\, + Q_{O3}=\, 0
\eeqa
we normalize charge such that a D3-brane in covering space has charge 
$+1$. With this convention an O3-plane has charge $-1/2$, and 
\beqa
N_{\rm flux}\, = \, \frac{1}{(4\pi^2\alpha')^2}\, \int_{X_6}\, H_3\wedge 
F_3 \,=\, \frac{1}{(4\pi^2\alpha')^2} \, \frac{i}{2\phi_I}\,
\int_{X_6} \, G_3\wedge {\ov G}_3
\eeqa
where $\phi_I$ is the imaginary part of the IIB complex coupling $\phi
=a+i/g_s$, and
\beqa
G_3=F_3-\phi H_3
\label{gflux}
\eeqa
Finally, in order to satisfy the equations of motion, the flux combination
$G_3$ must be imaginary self-dual with respect to the Hodge operation
defined in terms of the Calabi-Yau metric in $X_6$
\beqa
*_6 \, G_3\, = \,i\, G_3
\label{isd}
\eeqa
Given these conditions, the analysis in \cite{gkp} guarantees the 
existence of a consistent supergravity solution for the different relevant 
fields in the configuration, metric, and 4-form, which have the form of a 
warped compactification (similar to a black 3-brane solution, since the 
same fields are sourced).
\footnote{Interestingly, the above configurations lead, at the classical 
supergravity level, to 4d Poincare invariant 
solutions (i.e. vanishing cosmological constant) and flat potential for 
the overal Kahler parameter, even in the absence of supersymmetry
\cite{gkp}, although $\alpha'$ corrections \cite{bbhl} in general spoil 
this property. Even with supersymmetry, spacetime non-perturbative effects 
\cite{kklt} may generate supersymmetry breaking effects and generate a 
cosmological constant. We will have nothing to say about this familiar 
problem.}

We remark that the above condition should not be regarded as an additional 
constraint on the fluxes. Rather, for a set of fluxes in a fixed 
topological sector (i.e. in a fixed cohomology class), eq. (\ref{isd}) is 
a condition on the scalar moduli which determine the internal metric. The 
scalar potential is minimized at points in moduli space where (\ref{isd}) 
is satisfied, while fluxes induce a positive scalar potential at other 
points. Hence introduction of fluxes leads to a natural mechanism to 
stabilize moduli. Explicit expressions will be discussed later on, for the 
moment let us state the result in \cite{gkp,fp,kst} that generically all 
complex 
structure moduli and most Kahler moduli are stabilized by this mechanism.

\subsection{Supersymmetry}

The conditions for a configurations with 3-form fluxes to preserve some 
supersymmetry have been studied in \cite{gp}, and applied in explicit
constructions in \cite{fp,kst}. Let us review these results.

The 10d $\NN=2$ type IIB real supersymmetry transformation parameters
$\epsilon_L$, $\epsilon_R$, can be gathered into a complex one
$\epsilon=\epsilon_L+i\epsilon_R$. It is chiral in 10d, satisfying
$\Gamma_{10d}\epsilon=-\epsilon$, with $\Gamma_{10d}=\gamma^0\ldots
\gamma^9$. Compactification on $\IX_6$ splits this spinor with respect 
to $SO(6)\times SO(4)$ e.g. as
\beqa
\epsilon_L=\xi\otimes u+\xi^*\otimes u^*
\eeqa
where $\xi$ is a 6d chiral spinor $\Gamma_{6d}\xi=-\xi$, and $u$ is a 
4d chiral spinor $\Gamma_{4d}u=u$. For $\IX_6$ of generic $SU(3)$
holonomy only one component of $\xi$ is covariantly constant and provides
susy transformations in 4d.

On the other hand, the presence of the O3-planes and D3-branes in the
background preserves only those $\epsilon$ satisfying
\beqa
\epsilon_R=-\gamma^4 \ldots \gamma^9 \epsilon_L
\eeqa
Such spinors are of the form $\epsilon=2\xi\otimes u$.

The conditions for a flux to preserve a supersymmetry associated to a 
particular spinor component of $\xi$ are \cite{gp}
\beqa
G\xi=0 \quad ; \quad G\xi^*=0 \quad ; \quad G\gamma^m \xi^*=0
\label{susyspinor}
\eeqa
where $G=\frac 16 G_{mnl} \gamma^{[m}\gamma^n\gamma^{l]}$.

To understand this a bit better, let us introduce complex coordinates 
$z^i,{\ov z}^i$, where the gamma matrix algebra reads 
\beqa
\{\gamma^i ,\gamma^j\}=\{\gamma^{\ov i},\gamma^{\ov j}\}=0 \quad ; \quad
\{\gamma^i,\gamma^{\ov j}\}=\delta^{i{\ov j}}
\eeqa
Introducing the highest weight state $\xi_0$ satisfying $\gamma^{\ov 
i}\xi_0=0$, the spinor representation is

\begin{center}
\begin{tabular}{ccccc}
State & $SO(6)$ weight & \quad & State & $SO(6)$ weight\\
$\xi_0$ & $\frac 12(+++)$ & & $\gamma^i\xi_0$ & $\frac 12(\underline{-++})$ \\
$\gamma^1\gamma^2\gamma^3\xi_0$ & $\frac 12(---)$ &&
$\gamma^i\gamma^j\xi_0$ & $\frac 12(\underline{--+})$ 
\end{tabular}
\end{center}

The O3-planes preserve $\xi_0$ and $\gamma^{ij}\xi_0$. Of these, a general 
Calabi-Yau (on which $z_i$ are complex coordinates)
preserves only $\xi_0$, since it is $SU(3)$ invariant.

The conditions that a given flux preserves $\xi_0$, can be described 
geometrically \cite{gp,fp,kst} as

{\bf a)} $G_3$ is of type $(2,1)$ in the corresponding complex structure

{\bf b)} $G_3\wedge J=0$ where $J$ is the Kahler form

For explicit discussion of these conditions see below. Notice that a $G_3$ 
flux which is not $(2,1)$ in a complex structure, may still be 
supersymmetric if it preserves other spinor $\xi_0'$ (although it does not 
preserve $\xi_0$). In such case, $G_3$ would be of type $(2,1)$ in a
different complex structure where $\xi_0'$ is the spinor annihilated by
the new $\gamma^{{\ov i}{'}}$.

Since the techniques to find consistent (possibly supersymmetric) fluxes
at particular values of the stabilized moduli (and vice versa) have been 
discussed in the literature, we will not dwelve into their discussion. 

\medskip

In a generic Calabi-Yau compactification, where the holonomy is $SU(3)$
but not a subgroup of $SU(2)$), there is a unique component of
the spinor which is covariantly constant with respect to the spin 
connection. In this kind of situations, there is a preferred complex 
structure, that in which that spinor is annihilated by $\gamma^{\ov i}$. 
We will be interested in fluxes that preserve that spinor (so we call it 
$\xi_0$) and then $G_3$ should be $(2,1)$ in that complex structure.

In toroidal compactifications, the holonomy preserves several spinors, so 
we can play with different complex structures. In toroidal orbifolds, the 
orbifold projections in general project out some spinor components, 
leaving others invariant (just one if the orbifold group is in $SU(3)$ but 
not in $SU(2)$). In such case some complex structure is preferred, in 
analogy with the Calabi-Yau case.

\subsection{Fluxes in the $\IT^6/(\IZ_2\times \IZ_2)$}
\label{orbiflux}

We will be interested in discussing compactification on orbifolds with 
fluxes. The orbifold projection introduces some modifications with 
respect to the above toroidal case. 

i) Only some subset of fluxes is invariant under the orbifold projection.

ii) The orbifold space may contain closed cycles which are not closed in
the covering space. For instance, taking a $\IT^4$ parametrized by $x_i$
with identifications $x_i\simeq x_i+1$, modded out by $x_i\to -x_i$, one
such 2-cycle is provided by points $(x_1,x_2)$ with $0\leq x_1<1$,
$0\leq x_2< 1/2$. In general, the volume of such cycles is given by the
volume of some cycle in the covering space, divided by the order of the
orbifold. Hence if they exist they impose more restrictive quantization
constraints of the fluxes \footnote{This should be more properly 
understood as imposing quantization of the flux over a basis of cycles 
in the quotient space, whose elements in general involve linear 
combinations of untwisted and collapsed cycles. We thank R. Blumenhagen 
for discussion on this point.}.

iii) In addition, orbifolds contain new cycles, collapsed at the singular
points, which are not present in the parent torus. Thus one has to ensure
proper quantization of the field strength fluxes not just over the cycles
in the parent torus, but also over the cycles collapsed at the
singularities.

iv) Orbifolds contain additional moduli associated to the twisted 
sectors. In general, these scalar moduli also appear in the complete 
flux-induced scalar potential. In working at the orbifold limit, one
should make sure that this point in geometric moduli space indeed
corresponds to a minimum of the potential, with respect to these scalars.

v) As mentioned above, supersymmetry of the configuration requires the 
fluxes to preserve the spinor component which is invariant under the 
orbifold action.

\medskip

The systematic analysis of these features for arbitrary orbifolds seems 
rather difficult, so we prefer to center on a particular example.
We will be interested in the particular case of type IIB on 
$\IT^6/(\IZ_2\times \IZ_2)$ modded out by $\Omega R$. The geometry of the 
orbifold has been described in section \ref{geometry}.

Let us discuss the above issue in detail in the present setup.

iii) We consider fluxes only for untwisted NSNS and RR fields (constant
fluxes). This allows to apply the techniques developed for toroidal
orientifolds, and moreover ensures that their integral over 3-cyles
collapsed at the singular points is zero. The integral would be
proportional to the correlator $\langle {\cal O}_T {\cal O}_G\rangle$ of
the twisted form and 2-form vertex operators, which vanishes due to
mismatch of discrete charge under the $\IZ_2$ quantum symmetry. This
implies that these fluxes are properly quantized with respect to collapsed
cycles.

i) NSNS and RR 3-form fluxes invariant under the orbifold actions should 
involve one leg on each two torus. 

iv) We will ensure the fluxes are supersymmetric at the point in CY 
moduli space corresponding to the orbifold limit, so it is guaranteed that
the orbifold limit is a minimum of the scalar potential. In addition
for the $\IZ_2\times \IZ_2$ orbifold (with our choice of discrete torsion)
twisted sector moduli correspond to complex  structure deformations.
Following \cite{gkp}, we expect all these moduli to be stabilized (at the
orbifold point in moduli space) by the fluxes. Hence the orbifolds we
discuss do not contain additional moduli from twisted sectors.

v) There is a preferred spinor, surviving the orbifold projection (CY
holonomy), given by $\xi_0=1/2(+++)$. Hence to build susy fluxes 
we center on fluxes preserving the corresponding supersymmetry. Namely, of
$(2,1)$ kind in the complex structure given by $(z_1,z_2,z_3)$.

ii) Unfortunately, the $\IZ_2\times \IZ_2$ orbifold of ${\IT}^6/\Omega R$ 
contains a closed cycle, $0\leq x_i\leq 1/2$ for $i=4,6,8$, which is
not closed in ${\IT}^6/\Omega R$. This cycle requires the fluxes 
over toroidal cycles to be quantized to multiples of 8. The latter 
oversaturate the RR tadpole from the O3-planes and require the 
introduction of antibranes in the configuration, which is therefore 
non-supersymmetric. For simplicity we will preserve $\NN=1$ supersymmetry 
in the chiral sector of the theory, and simply add anti-D3-branes to 
cancel the excess of RR charge arising from the flux. 

\medskip

An additional advantage of our orbifold models is that the orbifold 
projection eliminates some of the (non-stabilized) Kahler moduli present 
in toroidal orientifolds. Hence, in a sense, our models stabilize (almost) 
all moduli by a combination of orbifold projections and NSNS and RR 
fluxes.

\section{Philosophy of our approach}
\label{philosophy}

The models considered in \cite{kst,fp} succeed in leading to $\NN=1$ 
or non-supersymmetric low-energy theories, with stabilization of 
most moduli. However, they are unrealistic in that they are automatically 
non-chiral, since the only gauge sectors live on parallel D3-branes whose 
low-energy spectra are non-chiral \footnote{The D3-brane world-volume
spectra are at best $\NN=1,0$ deformations of $\NN=4$ theories, by
flux-induced operators breaking partially or totally the
world-volume supersymmetry}.

We are interested in constructing models containing gauge sectors with 
charged chiral fermions, and with a bulk with flux-induced moduli 
stabilization. There are several possibilities to do this, corresponding 
to the different ways to build configurations of D-branes containing 
chiral fermions. In this paper we center on supersymmetric model building,
leaving several interesting non-supersymmetric setups for future work.

\medskip

{\bf D3-branes at singularities}

One possibility is to use compactification varieties containing singular 
points, e.g. orbifold singularities. Locating D3-branes at the singularity
leads to chiral gauge sectors, with low energy spectrum given by a quiver
diagram \cite{dm} \footnote{For model building in this setup with no 
field strength fluxes, see \cite{singu}.}. A simple possibility would be 
to consider
type IIB theory on $\IT^6/\IZ_3$ with the $\Omega R$ orientifold 
projection introduced above. This is particularly promising, since it is 
the simplest orbifold which can lead to three families in the sector of 
D3-branes at singularities. However, it is not possible to obtain $\NN=1$
supersymmetric models in this setup, for the following reason. In the
complex structure where the spinor invariant under $\IZ_3$ satisfies
$\gamma^{\ov i}\xi_0=0$, the $\IZ_3$ orbifold action reads
\beqa
\theta:(z_1,z_2,z_3)\to (e^{2\pi i/3} z_1, e^{2\pi i/3} z_2, e^{-4\pi 
i/3} z_3)
\eeqa
Fluxes preserving the same spinor $\xi_0$ should be of type $(2,1)$ in 
this
complex structure, namely linear combinations of 
\beqa
d{\ov z}^1dz^2dz^3 \quad ; \quad
dz^1d{\ov z}^2dz^3 \quad ; \quad
dz^1dz^2d{\ov z}^3 
\eeqa
Such fluxes are not invariant under the orbifold action, and cannot be
turned on. In other words, the only possible fluxes are not supersymmetric.
The same problem arises for other promising orientifolds, like
$\IZ_3\times \IZ_2\times \IZ_2$. So we will not pursue the construction
of $\NN=1$ susy models in the setup of D3-branes at singularities.

An alternative would be to give up supersymmetry, and build $\NN=0$ 
models of this kind. An amusing possibility is that of models where the
D-brane sector preserves some supersymmetry, while the closed string
sector is non-supersymmetric due to the combination of fluxes and
orbifold action. We provide an example of this kind in appendix
\ref{zthree}.

\medskip

{\bf Intersecting D6-branes}

Much progress has been made in D-brane model building using type IIA
D6-branes wrapped on intersecting 3-cycles in an internal space
(e.g.\cite{bgkl,afiru,bkl,rest,susy,susy2}) \footnote{See 
\cite{orbif} for early work leading to non-chiral models, and
\cite{reviews} for reviews}.
However it is difficult to introduce NSNS and RR field strength fluxes in
those setups; the difficulties can be seen from different perspectives.
The models usually contain O6-planes; this implies that to avoid the usual
no-go theorems about turning on fields strength fluxes in supergravity we
need combinations of fluxes which source the RR 7-form. A possible
combination is the RR 0-form field strength of type IIA (i.e. a
cosmological constant of massive type IIA) and the NSNS 3-form field
strength. This combination of fluxes has not been studied in the
literature, so it is not a convenient starting point.

One may think that T-dualizing three times a model with O3-planes and RR
and NSNS fluxes would yield the desired configurations with O6-planes.
However, T-duality acts in a very non-trivial way on $H_{NSNS}$,
transforming it into non-trivial components of the T-dual metric, which is
no longer Calabi-Yau \cite{halflat,torsion,kstt}. The final configuration 
indeed would
contain fluxes (RR and `metric fluxes') which source the RR 7-form, and
would lead to Poincare invariant 4d models (consistently with T-duality).
However, an analysis of the possible 3-cycles on which to wrap D6-branes in
the resulting complicated geometries is lacking.

\medskip

{\bf Magnetised D-branes}

We propose to take a different route. Although it is not often explicitly
stated, it is also possible to obtain chiral fermions from wrapped
D-branes, if the geometry of the wrapped manifold or the topology of the
internal world-volume gauge bundle are non-trivial. The chiral fermions
arise from the Kaluza-Klein reduction of the higher-dimensional
worldvolume fermions if the index of the corresponding Dirac operator is
non-zero.

We will center on a particular realization of this, corresponding to type
IIB compactified on $\IT^6$ (with an additional $\Omega R$ orientifold,
and possibly orbifold
projections), with configurations of D9-branes spanning all of spacetime,
namely wrapped on $\IT^6$ and with non-trivial gauge bundles on $\IT^6$.
The bundles are given by constant magnetic fields on each of the $\IT^2$,
so we call them magnetised D-brane configurations. They have been
considered in \cite{magnetised} in the absence of closed string field
strength fluxes.

In the absence of fluxes, these configurations are related by T-duality to 
configurations of intersecting D6-branes, so that any model of the latter 
kind can be translated very easily (see \cite{bgkl,rabadan}) to a 
magnetised  D9-brane setup. In particular, this shows that there exist 
$\NN=1$ supersymmetric models of magnetised D9-branes containing chiral 
fermions, by simply T-dualizing the models of intersecting D6-branes in 
\cite{susy}.

The advantage of using the magnetised D9-brane picture with O3-planes is 
that it is now straightforward to include NSNS and RR fluxes in the 
configuration, by applying the tools reviewed above (for the situation 
without D-branes). Notice that in the T-dual version of intersecting
branes this corresponds to turning on a complicated set of
NSNS, RR and metric fluxes. For the class of fluxes we consider, the
picture of magnetised D9-branes is more useful
\footnote{For the most general set of fluxes, the generalized mirror
symmetry in \cite{halflat} dictates the translation between configurations 
of A-type branes (D6-branes on 3-cycles) and B-type branes (D9-branes with
bundles). In this general situation both pictures are in principle equally
complicated.}.

In the following section we review magnetised D9-brane configurations 
without NSNS and RR fluxes. The new ingredients due to turning on the 
latter are described in section \ref{themodels}.

\section{Magnetised D-branes}
\label{magnetised}

In this section we review configurations of magnetised D9-branes before 
the introduction of fluxes. We first consider the case of toroidal
compactifications, and subsequently incorporate orientifold projections
and orbifold projections. The models, in the absence of bulk fluxes, are
T-dual to the models of intersecting D6-branes in toroidal
compactifications \cite{afiru}, toroidal orientifolds \cite{bgkl}
and orbifolds \cite{susy}.

\subsection{Magnetised D-branes in toroidal compactifications}
\label{toroidal}

We start with the simple case of toroidal compactification, with no 
orientifold projection. Consider the compactification of type IIB theory
on $\IT^6$, assumed factorizable \footnote{Without orbifold projections,
this requires a constrained choice of fluxes, stabilizing moduli at values
corresponding to a factorized geometry. In our orbifolds below, such moduli
are projected out by the orbifold, and hence are simply absent.} 

We consider sets of $N_a$ D9-branes, labelled D9$_a$-branes, wrapped 
$m_a^i$ times on the $i^{th}$ 2-torus $({\IT}^2)_i$ in $\IT^6$, and with 
$n_a^i$ units of magnetic flux on $(\IT^2)_i$. Namely, we turn on a 
world-volume magnetic field $F_a$ for the center of mass $U(1)_a$ gauge 
factor, such that
\beqa
m_a^i \, \frac 1{2\pi}\, \int_{\IT^2_{\,i}} F_a^i \, = \, n_a^i
\label{monopole}
\eeqa
Hence the topological information about the D-branes is encoded in the 
numbers $N_a$ and the pairs $(m_a^i,n_a^i)$ \footnote{Notice the change 
of roles of $n$ and $m$ as compared with other references. This however 
facilitates the translation of models in the literature to our language.} 

We can include other kinds of lower dimensional D-branes using this 
description. For instance, a D7-brane (denoted D7$_{(i)}$) sitting and a 
point in $\IT^2_{\,i}$ and wrapped on the two remaining two-tori
(with generic wrapping and magnetic flux quanta) is described by 
$(m^i,n^i)=(0,1)$ (and arbritrary $(m^j,n^j)$ for $j\neq i$); 
similarly, a D5-brane (denoted D5$_{(i)}$) wrapped on $\IT^2_{\,i}$ (with 
generic wrapping and magnetic flux quanta) and at a point in the remaining 
two 2-tori is described by $(m^j,n^j)=(0,1)$ for $j\neq i$; finally, a 
D3-brane sitting at a point in $\IT^6$ is described by $(m^i,n^i)=(0,1)$ 
for $i=1,2,3$. This is easily derived by noticing that the boundary 
conditions for an open string ending on a D-brane wrapped on a two-torus 
with magnetic flux become Dirichlet for (formally) infinite magnetic 
field.

\medskip

D9-branes with world-volume magnetic fluxes are sources for 
the RR even-degree forms, due to their worldvolume couplings 
\beqa
\int_{D9_a} C_{10} \quad ;\quad
\int_{D9_a} C_{8}\wedge \tr F_a \quad ;\quad
\int_{D9_a} C_{6}\wedge \tr F_a^{\, 2} \quad ;\quad
\int_{D9_a} C_{4}\wedge \tr F_a^{\, 3} 
\eeqa
Consistency of the configuration requires RR tadpoles to cancel. Following the discussion in \cite{afiru}, leads to the 
conditions
\beqa
&\sum_a N_a m_a^1 m_a^2 m_a^3 = 0 \nonumber \\
&\sum_a N_a m_a^1 m_a^2 n_a^3 = 0 & {\rm and}\; {\rm 
permutations} \; {\rm of} \; 1,2,3\nonumber \\
&\sum_a N_a m_a^1 n_a^2 n_a^3 = 0 & {\rm and}\; {\rm 
permutations} \; {\rm of} \; 1,2,3\nonumber \\
&\sum_a N_a n_a^1 n_a^2 n_a^3 = 0 
\eeqa
Which amounts to cancelling the D9-brane charge as well as the induced 
D7-, D5- and D3-brane charges.

Introducing for the $i^{th}$ 2-torus the even homology classes
$[{\bf 0}]_i$ and $[{\IT^2}]_i$ of the point and the
two-torus, the vector of RR charges of the one D9-brane in the $a^{th}$
stack is
\beqa
[{\bf Q}_a]\, =\, \prod_{i=1}^3\, (m_a^i [{\IT}^2]_i + n_a^i [{\bf 0}]_i) 
\label{charge}
\eeqa
The RR tadpole cancellation conditions read
\beqa
\sum_a \, N_a [{\bf Q}_a]\, = \, 0
\eeqa

The conditions that two sets of D9-branes with worldvolume magnetic 
fields
$F_a^i$, $F_b^i$ preserve some common supersymmetry can be derived from
\cite{bdl}. Indeed, it is possible to compute the spectrum of open strings
stretched between them and verify that it is supersymmetric if
\beqa
\Delta_{ab}^1 \pm \Delta_{ab}^2 \pm \Delta_{ab}^3 = 0
\eeqa
for some choice of signs. Here
\beqa
\Delta_i=\arctan\, [(F_a^i)^{-1}]-\arctan \,[(F_b^i)^{-1}]
\eeqa
and
\beqa
F_a^i=\frac{n_a^i}{m_a^i R_{x_i} R_{y_i}}
\eeqa
which follows from (\ref{monopole}).

\medskip

The spectrum of massless states is easy to obtain. The sector of open 
strings in the $aa$ sector leads to $U(N_a)$ gauge bosons and 
superpartners with respect to the 16 supersymmetries unbroken by the 
D-branes. In the $ab+ba$ sector, the spectrum is given by $I_{ab}$ chiral 
fermions in the representation $(N_a,{\ov N}_b)$, where 
\beqa
I_{ab}=[{\bf Q}_a]\cdot [{\bf Q}_b]=\prod_{i=1}^3 (n_a^im_b^i-m_a^i n_b^i)
\label{zeromodes}
\eeqa
is the intersection product of the charge classes, which on the basic 
classes $[{\bf 0}]_i$ and $[{\IT}^2]_i$ is given by the bilinear form
\beqa
\pmatrix{0 & -1 \cr 1 & 0}
\eeqa
The above multiplicity can be computed using the $\alpha'$-exact boundary
states for these D-branes \cite{bgkl}, or from T-duality with
configurations of intersecting D6-branes. We now provide an alternative
derivation which remains valid in more complicated situations where the
worldsheet theory is not exactly solvable. Consider for simplicity
a single two-torus. We consider two stacks of $N_a$ and $N_b$ branes
wrapped $m_a$ and $m_b$ times, and with $n_a$, $n_b$ monopole quanta.
Consider the regime where the two-torus is large, so that the magnetic 
fields are diluted and can be considered a small perturbation around the 
vacuum configuration. In the vacuum configuration, open strings within 
each stack lead to a gauge group $U(N_am_a)$ and $U(N_bm_b)$ respectively, 
which is subsequently broken down to $U(N_a)\times U(N_b)$ by the monopole 
background, via the branching
\beqa
U(N_a m_a)\times U(N_b m_b) \to U(N_a)^{m_a}\times U(N_b)^{m_b}\to 
U(N_a)\times U(N_b)
\label{branch1}
\eeqa
Open $ab$ strings lead to a chiral 10d fermion transforming in the 
bifundamental $(\fund_a,\antifund_b)$ of the original $U(N_a m_a)\times 
U(N_b m_b)$ group. Under the decomposition (\ref{branch1}) the 
representation splits as
\beqa
(\fund_a,\antifund_{\,b})\to ({\underline{\fund_a,\ldots}};{\underline{ 
\antifund_{\,b},\ldots}}) \to m_a m_b (\fund_a,\antifund_{\,b})
\label{branch2}
\eeqa
The 8d theory contains chiral fermions arising from these, because of the 
existence of a nonzero index for the internal Dirac operator (coupled to 
the magnetic field background). The index is given by the first Chern class
of the gauge bundle to which the corresponding fermions couples. Since it
has charges $(+1,-1)$ under the $a^{th}$ and $b^{th}$ $U(1)$'s, the
index is
\beqa
{\rm ind}\, \Dsl_{ab}\, = \, \int_{\IT^2}\, (F_a-F_b)= \frac{n_a}{m_a}
-\frac{n_b}{m_b}
\eeqa
Because of the branching (\ref{branch2}), a single zero mode of the Dirac
operator gives rise to $m_am_{\,b}$ 8d chiral fermions in the
$(\fund_a,\antifund_b)$ of $U(N_a)\times U(N_b)$. The number of chiral fermions
in the 8d theory in the representation $(\fund_a,\antifund_b)$ of the final
group is given by $m_a m_b$ times the index, namely
\beqa
I_{ab}= m_am_b\int_{\IT^2} (F_a -F_b) = n_a m_b-m_a n_b
\eeqa
The result (\ref{zeromodes}) is a simple generalization for the case of 
compactification on three two-tori.

An important property about these chiral fields is that they are localized 
at points in the internal space. From the string theory viewpoint this 
follows because boundary conditions for open strings with endpoints on 
D-branes with different magnetic fields require the absense of center of 
mass zero mode in the worldsheet mode expansion.
From the low energy effective theory viewpoint, this follows because such
strings behave as charged particles in a magnetic field. From elementary
quantum mechanics, such particles feel a harmonic oscillator potential and
are localized in the internal space. Excited states in the harmonic 
oscillator system (Landau levels) correspond to stringy oscillator 
(gonions \cite{afiru} in T-dual picture).

Notice that the field theory argument to obtain the spectrum is valid only
in the large volume limit. However, the chirality of the resulting
multiplets protects the result, which can therefore be extended to
arbitrarily small volumes. This kind of argument will be quite useful in
the more involved situation with closed string field strength fluxes, where
we do not have a stringy derivation of the results.

\subsection{Magnetised D-branes in toroidal orientifolds}
\label{orienti}

We are interested in adding orientifold planes into this picture, since
they are required to obtain supersymmetric fluxes. Consider type IIB on
$\IT^6$ (with zero NSNS B-field) modded out by $\Omega R$, with $R:x_m\to
-x_m$. This introduces 64 O3-planes, which we take to be all $O3^-$
(see section \ref{bfield} for subtleties). It also 
requires the D9-brane configuration to
be $\IZ_2$ invariant. Namely, for the $N_a$ D9$_a$-brane with topological
numbers $(m_a^i,n_a^i)$ we need to introduce their $N_a$ $\Omega R$ images
D9$_{a'}$ with numbers $(-m_a^i,n_a^i)$.

The RR tadpole cancellation conditions read
\beqa
\sum_a \, N_a [{\bf Q}_a]\, +\, \sum_a \, N_a [{\bf Q}_{a'}]\, - 32\,
[{\bf Q}_{O3}]\, =\, 0
\eeqa
with $[{\bf Q}_{O3}]= [{\bf 0}]_1\times [{\bf 0}]_2\times [{\bf 0}]_3$.
More explicitly
\beqa
&\sum_a N_a m_a^1 m_a^2 n_a^3 = 0 & {\rm and}\; {\rm 
permutations} \; {\rm of} \; 1,2,3\nonumber \\
&\sum_a N_a n_a^1 n_a^2 n_a^3 = 16 
\eeqa
Namely, cancellation of induced D7- and D3-brane charge. Notice that there 
is no net D9- or D5-brane charge, in agreement with the fact that the 
orientifold projection eliminates the corresponding RR fields 
\footnote{There is also an additional discrete constraint, previously 
unnoticed in the literature, which we would like to point out. It follows 
from a careful analysis of K-theory D-brane charge in the presence of 
orientifold planes. Following \cite{wittenkth}, the charge of D5-branes 
wrapped on some $\IT^2$ in the presence of O3-planes is classified by a 
real K-theory group which is $\IZ_2$. This statement is T-dual to the fact 
that D7-brane charge is $\IZ_2$ valued in type I theory. Following 
\cite{urangakth} RR tadpole cancellation requires cancellation of the 
K-theory D-brane charge. Hence the total induced D5-brane charge
on the D9$_a$-branes (without images) must be even in the
above configurations. This amounts to the condition
\beqa
\sum_a N_a m_a^1 n_a^2 n_a^3 = {\rm even} \; {\rm and}\; {\rm 
permutations} \; {\rm of} \; 1,2,3
\eeqa
The condition is non-trivial, and models satisfying RR tadpole conditions 
in homology, but violating RR tadpole conditions in K-theory can be 
constructed \cite{fernando}. Such models are inconsistent, as can be made 
manifest by introducing a D7-brane probe, on which world-volume the 
inconsistency manifests as a global gauge anomaly \cite{urangakth}.
The condition is however happily satisfied by models in the literature, 
and also in our examples below.}

The rules to obtain the spectrum are similar to the above ones, with the 
additional requirement of imposing the $\Omega R$ projections. This 
requires a precise knowledge of the $\Omega R$ action of the different 
zero mode sectors (in field theory language, on the harmonic oscillator
groundstates for chiral fermions). The analysis is simplest in terms of the
T-dual description, where it amounts to the geometric action of the
orientifold on the intersection points of the D-branes. The result, which
is in any case derivable in our magnetised brane picture, can be taken from
\cite{bgkl}.

The $aa$ sector is mapped to the $a'a'$ sector, hence suffers no 
projection \footnote{We do not consider branes for which $a=a'$ here; 
they will be taken care of explicitly in the examples below.}. We obtain 
a 4d $U(N_a)$ gauge group, and superpartners with respect to the $\NN=4$ 
supersymmetry unbroken by the brane.

The $ab+ba$ sector is mapped to the $b'a'+a'b'$ sector, hence does not 
suffer a projection. We obtain $I_{ab}$ 4d chiral fermions in the 
representation $(\fund_a,\antifund_b)$. Plus additional scalars which are 
massless in the susy case, and tachyonic or massive otherwise.

The $ab'+b'a$ sector is mapped to the $ba'+a'b$. It leads to $I_{ab'}$ 4d 
chiral fermions in the representation $(\fund_a,\fund_b)$ (plus additional 
scalars).

The $aa'+a'a$ sector is invariant under $\Omega R$, so suffers a 
projection. The result is $n_{\Yasymm}$ and $n_{\Ysymm}$ 4d chiral
fermions in the $\Yasymm_a$, $\Ysymm_a$ representations, resp, with
\beqa
n_{\Yasymm}= \frac 12(I_{aa'}+ 8I_{a,O3})=\,- 4 \,m_a^1 m_a^2 m_a^3 \,
(n_a^1n_a^2n_a^3+1) \nonumber \\
n_{\Ysymm}= \frac 12(I_{aa'}-8I_{a,O3})=\,- 4 \,m_a^1 m_a^2 m_a^3 \,
(n_a^1n_a^2n_a^3-1) 
\eeqa
where $I_{a,O3}=[{\bf Q}_a]\cdot [{\bf Q}_{O3}]$.

\subsection{Magnetised D-branes in the $\IT^6/(\IZ_2\times\IZ_2)$ orbifold}

Finally, we will be interested in models with orbifold and orientifold
actions. In particular, consider type IIB on the orbifold
$\IT^6/(\IZ_2\times \IZ_2)$ described in section \ref{geometry}, modded out
by $\Omega R$. The model contains (for zero B-field, see section 
\ref{bfield}) 64 O3-planes (with $-1/2$ units of
D3-brane charge), and 4 O7$_i$-planes (with $-8$ units of D7$_i$-brane
charge), transverse to the $i^{th}$ two-torus. Their total charges are
given by $-32$ times the classes
\beqa
& [{\bf Q}_{O3}]=\, [{\bf 0}_1]\times [{\bf 0}_2]\times [{\bf 0}_3] \quad 
; \quad &
 [{\bf Q}_{O7_1}]=\,-\,[{\bf 0}_1]\times [({\IT}^2)_2]\times 
[({\IT}^2)_3] 
\nonumber \\
& [{\bf Q}_{O7_2}]=\,-\,[({\IT}^2)_1]\times [{\bf 0}_2]\times 
[({\IT}^2)_3] \quad ; 
\quad &
 [{\bf Q}_{O7_3}]=\,-\,[({\IT}^2)_1]\times [({\IT}^2)_2]\times [{\bf 
0}_3]
\label{charges}
\eeqa
where the signs are related to the specific signs in the definition of the 
$\IZ_2\times \IZ_2$ action.
We define $[{\bf Q}_{Op}]= [{\bf Q}_{O3}]+[{\bf Q}_{O7_1}]+
[{\bf Q}_{O7_2}]+[{\bf Q}_{O7_3}]$. The RR charge is cancelled using
magnetised D9-branes and their orientifold images (the orbifold projection
maps each stack of D9-branes to itself), which carry just induced D7$_i$-
and D3-brane charges. The RR tadpole conditions read
\beqa
\sum_a \, N_a [{\bf Q}_a]\, +\, \sum_a \, N_a [{\bf Q}_{a'}]\, - 32\,
[{\bf Q}_{Op}]\, =\, 0
\eeqa
The models with magnetised  D9-branes in this
orientifold are T-dual to those in \cite{susy}, whose main features are
easily translated. The spectrum can be computed using the above techniques,
taking care of the additional orbifold projections on the spectrum, or
equivalently translated from \cite{susy}. The result is shown in table
\ref{matter}, where $I_{a,Op}=[Q_a]\cdot [Q_{Op}]$.

\begin{table}[htb] \footnotesize
\renewcommand{\arraystretch}{1.25}
\begin{center}
\begin{tabular}{|c|c|}
\hline
\hspace{2cm} {\bf Sector} \hspace{2cm} &
\hspace{1.5cm} {\bf Representation} \hspace{2.5cm} \\
\hline\hline
$aa$   &  $U(N_a/2)$ vector multiplet  \\
       & 3 Adj. chiral multiplets   \\
\hline\hline
$ab+ba$   & $I_{ab}$ $(\fund_a,\antifund_b)$ fermions   \\
\hline\hline
$ab'+b'a$ & $I_{ab'}$ $(\fund_a,\fund_b)$ fermions   \\
\hline\hline
$aa'+a'a$ & $\frac 12 (I_{aa'} - 4 I_{a,Op}) \;\;
\Ysymm\;\;$ fermions  \\
          & $\frac 12 (I_{aa'} +  4 I_{a,Op}) \;\;
\Yasymm\;\;$ fermions \\
\hline
\end{tabular}
\end{center}
\caption{\small General chiral spectrum on generic magnetised 
D9$_a$-branes in the $\Omega R$ orientifold of $\IT^6/(\IZ_2\times \IZ_2)$. 
The models may contain additional non-chiral pieces
which we ingnore here. In supersymmetric situations, scalars combine
with the fermions given above to form chiral supermultiplets.
\label{matter} }
\end{table}           

\section{The models}                
\label{themodels}

\subsection{Construction of the models}

The models we are going to construct have the following structure. We 
consider type IIB theory on $\IT^6/(\IZ_2\times \IZ_2)$, where the 
orbifold twists act as 
\beqa
\theta & : & (z_1,z_2,z_3)\to (-z_1,-z_2,z_3) \nonumber\\
\omega & : & (z_1,z_2,z_3)\to (z_1,-z_2,-z_3) 
\eeqa
and mod out by $\Omega R$, where 
\beqa
R & : & (z_1,z_2,z_3)\to (-z_1,-z_2,-z_3) 
\eeqa
As explained above, the model contains \footnote{Assuming no discrete 
B-field, see section \ref{bfield}.} 64 O3-planes and 4 O7$_i$-planes 
(localized at 
points in the $i^{th}$ two-torus, and spanning the remaining two).
The orbifold and orientifold projections preserve the spinor satisfying 
$\gamma^{\ov i}\xi=0$ in the above complex coordinates.

We include a combination of NSNS and RR 3-form field strength fluxes
invariant under the orbifold/orientifold actions, and which preserve the
same spinor $\xi_0$. This means that the corresponding 3-form $G_3$ is of
type $(2,1)$ in the above complex coordinates, and involves all three
complex coordinates. The general form of such flux is
\footnote{In what follows we absorb the normalization factor $1/(4\pi^2
\alpha')$ in the definition of the fluxes.}
\beqa
G_3= g_1\, d{\ov z}_1dz_2dz_3\, +\, g_2\, dz_1d{\ov z}_2dz_3\, +\,
g_3\, dz_1dz_2d{\ov z}_3
\eeqa
To verify that the flux is supersymmetric, it must also satisfy the
primitivity condition $G_3\wedge J =0$. In fact, the general Kahler form of
$\IT^6/(\IZ_2\times \IZ_2)$ is factorizable
\beqa
J=J_1 dz_1d{\ov z}_1+J_2dz_2d{\ov z}_2+J_3 dz_3d{\ov z}_3 
\eeqa
since the off-diagonal pieces of the metric are projected out by the
orbifold action. Hence the above flux is automatically primitive and thus
supersymmetric. This also implies that the Kahler parameters of the
three two-tori are not stabilized and have flat potential.

As discussed quite generally in \cite{gkp,fp,kst}, the flux stabilizes 
most geometric moduli. This follows because fluxes are quantized, hence 
the local values of the field strengths depend on moduli (controlling the 
sizes of cycles). Hence only for particular values of the moduli the 
minimization of energy conditions are obeyed by the local values of the 
field strengths. In our expression above, the stabilization of moduli 
arises in a formally different way: we have already ensured that the local 
field strength minimize the energy (and in fact determine a supersymmetric 
vacuum), but only for very particular values of the moduli the fluxes turn 
out to be quantized. These are the values for which the moduli are
stabilized by the fluxes.

The conditions of proper quantization are easy to obtain (but in general
difficult to solve systematically). The periods of $dz_i$, $d{\ov z}_i$ over
the basis 1-cycles $[a_i]$, $[b_i]$, in the two-tori are
\beqa
\int_{[a_i]}dz_j=\delta_{ij} \quad ; \quad \int_{[b_i]} dz_j=\tau_i 
\delta_{ij} \; ({\rm no}\; {\rm sum})
\eeqa
We also have
\beqa
F_3=-\frac{{\ov \phi} G_3 - \phi {\ov G}_3}{\phi-{\ov \phi}} \quad ; \quad 
H_3=-\frac{G_3-{\ov G}_3}{\phi-{\ov \phi}}
\eeqa
Hence, the proper quantization of $F_3$, $H_3$ on the different 3-cycles 
requires 
\beqa
& \int_{[a_1]\times [a_2]\times [a_3]} F_3 \in 8\times {\IZ} \quad ; 
\quad
\int_{[a_1]\times [a_2]\times [b_3]} F_3 \in 8\times {\IZ} \quad {\rm 
and}\; {\rm perms.} \nonumber \\
& \int_{[b_1]\times [b_2]\times [b_3]} F_3 \in 8\times {\IZ} \quad ; 
\quad
\int_{[b_1]\times [b_2]\times [a_3]} F_3 \in 8\times {\IZ}
\quad {\rm and}\; {\rm perms.}
\eeqa
and similarly for $H_3$.
This determines the values at which untwisted complex structure moduli 
stabilize. The equations are difficult to solve in general, and different
tecniques to construct consistent supersymmetric fluxes have been described
in the literature \cite{kst}. In what follows we will simply choose some
particular solutions interesting for our applications. 

On the other hand, the fluxes were known from the start to be properly 
quantized on twisted 3-cycles, when twisted moduli sit at the orbifold 
point in moduli space. This guarantees that the point in moduli space
corresponding to the orbifold geometry is a minimum of the twisted
scalar potential. From general analysis \cite{gkp}, we moreover expect that
these additional moduli are stabilized at values corresponding to the
orbifold geometry, although it would be more involved to verify
this more explicitly (in particular it would require an expression of 
properly quantized fluxes as a funcion of the complex structure 
deformation of the singularities).

Therefore, the only moduli remaining in the configuration are the Kahler 
parameters for each of the two-tori. In our models below we will 
make special choices which make the chiral sector of D9-brane 
configurations 
supersymmetric (although this does not imply that these parameters are
stabilized \cite{susy}).

The contribution from such flux to the $C_4$ RR tadpole is
\beqa
N_{\rm flux}\, =\, \frac{4\tau_{1,I}\tau_{2,I}\tau_{3,I}}{\phi_I}
\, \left( |g_1|^2+|g_2|^2+|g_3|^2 \right)
\label{nflux}
\eeqa
The remaining RR tapdole for $C_4$ as well as for the RR 8-forms, must 
be cancelled by the addition of D-branes. We introduce a set of $N_a$ 
D9$_a$-branes, with wrapping numbers and world-volume magnetic fluxes 
given by $(m_a^i,n_a^i)$. Their RR charges are encoded in an even 
homology class $[{\bf Q}]$ of the form (\ref{charge}). For each O-plane
and the flux, we have the charges (\ref{charges}) and
\beqa
& [{\bf Q}_{\rm flux}]= \,N_{\rm flux} \,
[{\bf 0}_1]\times [{\bf 0}_2]\times [{\bf 0}_3]
\eeqa
The RR tadpole cancellation conditions for them read
\beqa
[{\bf Q}_{\rm tot.}]\, =\, \sum_a N_a [{\bf Q}_a]\, +\,
\sum_{a'} N_a [{\bf Q}_{a'}]\, +\, [{\bf Q}_{\rm flux}]\, 
-32\, [{\bf Q}_{Op}]\, =\, 0
\eeqa
or more explicitly
\beqa
&\sum_a N_a n_a^1 n_a^2 n_a^3 + \frac 12 N_{\rm flux} -16= 0 \nonumber\\
&\sum_a N_a n_a^1 m_a^2 m_a^3 +16 = 0 \nonumber\\
&\sum_a N_a m_a^1 n_a^2 m_a^3 +16 = 0 \nonumber\\
&\sum_a N_a m_a^1 m_a^2 n_a^3 +16 = 0 
\label{rrtadpoles}
\eeqa
Once these conditions are satisfied, a consistent model results. 

\medskip

We will also require that the chiral part of the D-brane configuration 
preserves the $\NN=1$ 
supersymmetry unbroken by the orbifold/orientifold, and by the fluxes. 
The condition on the worldvolume magnetic fields is
\beqa
\sum_i \arctan \frac{m_a^i A_i}{n_a^i}=0
\eeqa
When taken for a fixed set of integers $(m_a^i,n_a^i)$, it turns into a
constraint on the Kahler moduli of the orbifold. For a small number of
D-brane stacks, they can be satisfied by simply adjusting the 
Kahler parameters of the three two-tori, which are not fixed by the
flux-induced potential, as discussed above. For larger numbers, the
conditions typically overconstrain the Kahler parameters and the
corresponding D-brane configuration cannot be made supersymmteric
\footnote{Notice that solving for some Kahler parameters does not
imply they are stabilized. As discussed in \cite{susy}, it only implies 
that for other choices of Kahler moduli the ansatz for the D-brane
configuration is non-supersymmetric, and decays into a stable and
supersymmetric one. In the T-dual picture it corresponds to recombining
intersecting D-branes; in the magnetised D-brane picture it corresponds to
transitions to configurations where the gauge bundle involves the
non-abelian pieces.}.

Two subtle points arise in the above construction, related to some 
potential peculiar features of D-branes (concretely, the K-theory group 
classifying D-brane charges) in the presence of NSNS 3-form 
fields. They are discussed in the appendix \ref{ktheory}, with the outcome 
that their presence can be ignored in the following discussion.

\subsection{The spectrum}

Let us describe the corresponding spectrum. Clearly, the presence of the 
NSNS and RR field strength fluxes does not allow for an $\alpha'$-exact 
quantization of the 2d worldsheet theory. Therefore the spectrum can only 
be described in the supergravity approximation, namely for large internal 
volumes. In this regime, the fluxes are dilute and their effect is small 
and under control. For some quantities, protected by supersymmetry and/or 
chirality, we are allowed to extrapolate the large volume result to the 
small volume regimes
\footnote{More explicitly, starting from a configuration which is stable 
at some value of the moduli, we extrapolate it to large volume (keeping 
the topological classes of each D-brane fixed, hence allowing for unstable 
branes in the large volume limit). There the chiral spectrum is computed, 
and is extrapolated back to the stable point in moduli space. 
Consideration of unstable configurations at intermediate steps is not 
problematic, since we are interested in topological quantities.}.

The discussion of the closed string sector is as in models without 
D-branes \cite{gkp,fp,kst}. We obtain the 4d $\NN=1$ supergravity 
multiplet, and three neutral chiral multiplets associated to the Kahler 
parameters of the three two-tori. On top of these massless states, we have 
a tower of states with masses induced by the NSNS and RR fluxes, of order 
$\alpha'/R^3$ for large volume (much lighter than the KK replicas).

We are more interested in determining the spectrum on the new sector, the 
D-branes, and whether it is modified by the NSNS and RR fluxes. Since the 
fluxes are `color blind', in the sense that they can couple only to traces 
over the gauge indices, each stack of $N_a$ D9$_a$-branes still leads to
a $U(N_a)$ gauge symmetry. By supersymmetry we obtain 4d $\NN=1$ vector 
multiplets with gauge group $\prod_a U(N_a)$, just like in the absence of 
bulk fluxes. This large volume result is presumably robust enough to be 
maintained at small volumes.

There is however a non-trivial effect from the fluxes. The usual three 
adjoint chiral multiplets in the $aa$ open string sector are in general 
expected not to be present, due to interactions with the closed 
sector, which is only $\NN=1$ supersymmetric. The masses for these chiral 
multiplets seem not to arise at leading order in $\alpha'$ \cite{kst}, but 
are certainly expected in the full theory. This effect has the additional 
advantage of removing the corresponding moduli fields in the open string 
sector. We also expect this to hold  for any non-chiral set of fields not 
protected by supersymmetry, chirality or gauge symmetry. 

In $ab$ sectors (or similarly $ab'$ or $aa'$ sectors), the massless modes 
are given by $\NN=1$ chiral multiplets transforming as bifundamental 
representations of the corresponding gauge groups. These multiplets arise 
from zero modes of the Dirac operator coupled to the D9-brane world-volume 
magnetic field, and are localized at points in the internal space. This 
implies that in the large volume limit these fields are insensitive to the 
presence of NSNS and RR field strengths, since their density is small at 
the location of the zero mode. That is, the turning on of fluxes can be 
regarded as an adiabatic change in the local region around the zero modes, 
hence the index of the Dirac operator is unchanged by them. The massless 
spectrum is therefore still given by the same formulas as in the situation 
without fluxes. By chirality, the same result holds in the small volume 
regime. Note that the spectrum of massive excitations (higher Landau 
levels) will in general be modified by the bulk fluxes.

The main conclusion is that the spectrum of gauge multiplets and chiral 
matter is still given by the above formulae in table \ref{matter} for our 
models in 
$\IT^6/(\IZ_2\times \IZ_2)$ \footnote{In non-susy cases of toroidal 
orientifolds with fluxes and D-branes, we also expect the spectrum 
to be of the familiar form (reviewed in section \ref{toroidal}) to be 
valid at large volume, and by chirality at smaller volume for gauge 
fields and chiral fermions.}. We will find additional
support for this coming from the analysis of anomalies. Masses of 
vector-like fields will in general change, and it would be interesting to 
have an estimate for them. 

\subsection{Anomaly cancellation}

{\bf Cubic non-abelian anomalies}

There is a close relation between cancellation of RR tadpoles and 
cancellation of anomalies. In models with NSNS and RR 
fluxes, where there 
is an additional contribution to the RR tadpole conditions, it is expected 
that the usual anomaly cancellation pattern is different. This issue was 
analyzed in \cite{urangaflux}, where it was determined that the presence 
of NSNS and RR fluxes induce explicitly non-gauge invariant 
Wess-Zumino terms in the action of D-branes.

In our above setup, the coupling on the volume of the D9$_a$-brane 
stack relevant for the discussion of 4d anomalies is of the form
\beqa
\int_{D9_a} B_{NS}\wedge F_3 \wedge (\tr F_a^3)^{(0)} 
\eeqa
where we are using Wess-Zumino descent notation \footnote{Namely for any 
closed gauge invariant form $Y$ we define $Y=dY^{(0)}$ and $\delta 
Y^{(0)}=dY^{(1)}$, with $\delta$ denoting gauge variation.}. The 
variation of this term under a $SU(N_a)$ gauge transformation induces a 4d 
anomaly given by
\beqa
& \int_{D9_a} B_{NS}\wedge F_3 \wedge \delta(\tr F_a^3)^{(0)} =
\int_{D9_a} B_{NS}\wedge F_3 \wedge d(\tr F_a^3)^{(1)} = \nonumber \\
& =\int_{D9_a} H_3\wedge F_3 \wedge (\tr F_a^3)^{(1)} =
m_a^1m_a^2m_a^3\, N_{\rm flux}\,\int_{M_4} \, (\tr F_a^3)^{(1)} =
& [Q_a]\cdot [Q_{\rm flux}] \, \int_{M_4} \, (\tr F_a^3)^{(1)}
\nonumber
\eeqa
Further quotient by the orientifold action cut this contribution by one 
half.

The total $SU(N_a/2)^3$ anomaly is given by this contribution plus the 
familiar field theory triangle diagrams where dynamical fermions in the 
spectrum run in a loop. Their net contribution is given by
\beqa
&\sum_{b \ne a}^{} I_{ab} N_b/2 + \sum_{b'\ne a'}^{} I_{ab'} N_b/2 + n_{\Yasymm} (N_a/2-4)
+ n_{\Ysymm} (N_a/2+4)=& \nonumber \\
&= [Q_a]\cdot \frac 12 (\sum_{\forall b}^{} N_b [Q_b] + \sum_{\forall b}^{} 
N_{b}[Q_{b'}] -32 [Q_{Op}]) &
\eeqa
Thus the total contribution is proportional to $[Q_a]\cdot [Q_{\rm tot.}]$
which vanishes due to the RR tadpole condition (\ref{rrtadpoles}).

It is easy to check that other anomalies, like mixed $U(1)$-$SU(N_a)^2$ or 
mixed gravitational anomalies also cancel, involving in addition a 
Green-Schwarz mechanism identical to that in models without NSNS and RR 
fluxes \cite{afiru,susy}.

\medskip

{\bf Brane-bulk mixed anomalies}

In the presence of fluxes the 10d supergravity Chern-Simons interaction 
(\ref{cs}) can lead to the mixing of 4d 1-form $U(1)$ gauge fields (from 
the KK reduction of the RR-form along a 3-cycle) with 4d scalar (4d dual 
to the NSNS or RR 2-form) \cite{strongcp}. Indeed this kind of coupling is 
responsible for giving masses to 4d gauge bosons which are superpartners 
of the stabilized moduli \cite{fp,kst}.

On the other hand, this kind of $B\wedge F$ couplings
can lead to $U(1)$-$SU(N_a)^2$ mixed anomalies between a closed string 
$U(1)$ gauge field and the $SU(N_a)$ gauge bosons on D9-brane 
world-volumes, by a Green-Schwarz type diagram. Such anomalies would be 
however canceled by flux-induced Wess-Zumino terms in the D9-brane 
world-volumes \cite{strongcp}.

This phenomenon, although present in general models with field strength 
fluxes, is absent in our configurations. This is because the closed string 
sector does not contain any gauge bosons; equivalently, because 
e.g. components of the RR 4-form with three indices in the internal 
directions are odd under the $\Omega R$ projection. Hence our models do 
not have Green-Schwarz diagrams contribution to mixed anomalies 
involving bulk and brane gauge interactions.

\subsection{Examples}

{\bf A simple chiral model}

Let us start by considering a simple and illustrative example of a chiral
model. Consider the flux
\beqa
G_3\, =\, 4\times\, \frac{2}{\sqrt{3}}\, e^{-\pi i/6}\, (\, d{\ov 
z}_1dz_2dz_3\, +
\, dz_1d{\ov z}_2dz_3\, +\, dz_1dz_2d{\ov z}_3\, )
\eeqa
which corresponds to the particular example of fluxes (4.18) in \cite{kst}.
The flux stabilizes the complex structure and complex coupling moduli at
values
\beqa
\tau_1=\tau_2=\tau_3=\phi=e^{2\pi i/3}
\eeqa
The flux is supersymmetric, and it is easy to check that its integral over
any 3-cycle of $\IT^6$ is multiple of 8

The contribution of the flux to the 4-form RR tadpole (\ref{nflux}) is
\beqa
N_{\rm flux}=192
\eeqa
We can satisfy the RR tadpole conditions (\ref{rrtadpoles}) with the 
set of magnetised D9-branes in table \ref{chiralone} (and their 
$\Omega R$ images).
The main part of the model (except for the 180 ${\ov D3}$-branes in the 
last line) is supersymmetric for $\arctan A_1+\arctan A_2-\arctan A_3=0$

\begin{table}[htb] \footnotesize
\renewcommand{\arraystretch}{1.25}
\begin{center}
\begin{tabular}{|c||c|c|c|}
\hline
$N_a$ & $(m_a^1,n_a^1)$ & $(m_a^2,n_a^2)$ & $(m_a^3,n_a^3)$ \\
\hline\hline
 $10$ & $(1,1)$ & $(1,1)$ & $(-1,1)$ \\
 $6$ & $(0,1)$ & $(1,0)$ & $(-1,0)$ \\
 $6$ & $(1,0)$ & $(0,1)$ & $(-1,0)$ \\
 $26$ & $(1,0)$ & $(-1,0)$ & $(0,1)$\\
$90$ & $(0,-1)$ & $(0,-1)$ & $(0,-1)$\\
\hline
\end{tabular}
\end{center}
\caption{\small D9-brane configuration for the $SU(5)$ GUT model}
\label{chiralone}
\end{table}   

The D9-branes along classes invariant under $\Omega R$ (refered to as
filler branes) lead with vanishing Wilson lines to a maximally enhanced
gauge group
$USp(6)\times USp(6)\times USp(26)$, and $\NN=1$ chiral multiplets
in three copies of the two-index antisymmetric representation of the
corresponding symplectic factor. We also get $U(1)$ factors for the 
${D3}$-branes, taken non-coincident. The remaining stack of 10 D9-branes
(referred to as GUT branes),
and their $\Omega R$ images, lead to a $U(5)$ $\NN=1$ vector multiplets
and three adjoint chiral multiplets. Additional non-chiral matter
arises from strings stretching among the filler branes. It is expected
that $\alpha'$ effects generate tree-level masses for all these non-chiral
matter multiplets. The interesting chiral matter arises from open strings
stretching between the GUT branes and their images, and between the GUT
branes and the filler branes. In total, we have the matter content
\beqa
& U(5)\times USp(6)\times USp(6)'\times USp(26)& \nonumber \\
& 8(10;1,1,1) + (5;6,1,1) + (5;1,6,1) + ({\ov 5};1,1,26) + 
90({\ov 5};1,1,1)&
\eeqa
The spectrum is anomalous, but the anomaly is exactly cancelled by the
flux induced 4d Wess-Zumino terms, which can be checked to contribute
as 96 chiral fermions in the fundamental representation of $U(5)$.

The above example contains a subsector providing a peculiar kind of
$SU(5)$ grand unified theory, with 8 families. It is peculiar in that
it contains additional fundamental representations unmatched with
antifundamentals, so that Dirac mass terms are in principle not possible
to get rid of the additional chiral matter, since its anomaly is cancelled
via Wess-Zumino terms. 

As discussed in section \ref{extrabranes}, the model contains some 
additional D-branes, which are required for consistency of the stacks of 
D9-branes wrapped on cycles with non-zero $H_3$. Their presence, however, 
does not introduce any additional interesting features in the model (at 
least, concerning its supersymmetry and its chiral spectrum).

\medskip

{\bf Another more interesting chiral model}

The existence of Wess-Zumino terms for gauge factors in the would-be
Standard Model is not in principle phenomenologically desirable. After 
all, the gauge representations of standard model fermion {\em are} 
anomaly-free.
This is not a general obstacle for model building with D-branes in the
presence of fluxes. Indeed it is possible to construct models where the
would-be Standard Model sector has no Wess-Zumino terms, because the
corresponding D-branes have classes with zero intersection with the
class of the flux. In this kind of models, the visible sector also does 
not intersect the ${\ov D3}$-branes, and is hence fully supersymmetric, 
with supersymmetry broken in a hidden sector.

Consider the same flux configuration as above, but now let us satisfy
RR tadpole cancellation using the stacks of D-branes in table 
\ref{chiraltwo}, plus 180 ${\ov D3}$-branes. The visible sector of this 
configuration is supersymmetric for $A_1=2A_2=A_3$. Notice that 
since every stack of branes has at least some $m_a^i=0$, they are really 
D7-branes wrapped on two complex planes, with non-trivial worldvolume 
magnetic fields along them.

\begin{table}[htb] \footnotesize
\renewcommand{\arraystretch}{1.25}
\begin{center}
\begin{tabular}{|c||c|c|c|}
\hline
$N_a$ & $(m_a^1,n_a^1)$ & $(m_a^2,n_a^2)$ & $(m_a^3,n_a^3)$ \\
\hline\hline
$6$ & $(1,1)$ & $(-2,1)$ & $(0,1)$ \\
$4$ & $(0,1)$ & $(2,1)$ & $(-1,1)$ \\
$4$ & $(1,0)$ & $(-1,0)$ & $(0,1)$ \\
$16$ & $(1,0)$ & $(0,1)$ & $(-1,0)$ \\
$8$ & $(0,1)$ & $(1,0)$ & $(-1,0)$ \\
\hline
\end{tabular}
\end{center}
\caption{\small D-brane configuration for the four-family model}
\label{chiraltwo}
\end{table}

The resulting spectrum of vector and chiral multiplets is (at the chiral
level)
\beqa
& U(3)\times U(2)\times USp(4)\times USp(16)\times USp(8) &\nonumber \\
& 4(3,{\ov 2};1,1,1)_{1,-1} + 2({\ov 3},1;1,16,1)_{-1,0} +
(3,1;1,1,8)_{1,0} + (1,{\ov 2};4,1,1)_{0,-1}
+ & \nonumber \\
& + 2(1,2;1,16,1)_{0,1} + 2(3,1;1,1,1)_{-2,0} + 2(6,1;1,1,1)_{2,0} +
2(1,1;1,1,1)_{0,2} + 2(1,3;1,1,1)_{0,-2} & \nonumber
\eeqa
The model is very close to a four-family Standard Model (in fact, appeared
in \cite{susy}, with some branes replaced by fluxes). However, it lacks
a massless candidate $U(1)$ for hypercharge, since the suitable linear
combination become massive due to $B\wedge F$ couplings \cite{imr}. Hence
the model is admittedly not realistic, but it provides a good illustration
of model building possibilities. Notice that the supersymmetric visible 
sector is really  decoupled (modulo closed string interactions) from the 
supersymmetry breaking hidden sector. Note also that the latter is 
relatively stable, since annihilation processes between fluxes and branes 
are quite suppressed \cite{klst}, and is in fact very similar to recent 
proposal to build de Sitter vacua in string theory with large flux quanta 
and ${\ov D3}$-branes \cite{kklt}. 

It is not our purpose to start a systematic exploration of model building
in this setup, which we leave for future work. We would like to
emphasize however that much more freedom is expected if one considers
the possibility of taking flat directions in initial supersymmetric
models with larger gauge group. This has been successfully exploited in 
other supersymmetric models in the literature. We leave this avenue for 
richer model building for future work.

\section{Discrete B field}
\label{bfield}

The configurations we are considering admit a discrete modification, 
corresponding to including a discrete value of the NSNS 2-form field along 
some of the two-tori \cite{bkl}. This possibility is interesting as a 
possible way of obtaining models with odd number of fermion families 
(although not the only one, if $SU(2)_L$ is embedded as $USp(2)$ in some 
filler branes \cite{yuk}). In this section we describe the main 
modifications introduced by non-trivial B-fluxes, centering for simplicity 
on the case of B-flux on a single two torus, say the third.

A first modification \cite{bkl} is that D9-branes carry induced 
lower-dimensional brane charges due to the $1/2$ unit of $B_{NSNS}$. 
It is then useful to introduce an effective world-volume magnetic flux 
quantum, defined by ${\tilde n}_a^3=n_a^3+\frac 12 m_a^3$.

A second effect is that the discrete B-field changes the RR tadpole 
cancellation conditions, decreasing by a factor of two the total charge 
associated to O-planes pointlike in the corresponding two-torus. The 
physical interpretation of this has been discussed quite explicitly in
section 3 of \cite{witvec}: $1/4$ of the corresponding O-planes 
carry positive RR charge, while $3/4$ of them are the familiar 
negatively charged O-planes. The net effect is to cut by half the total 
contribution of the O-planes to the RR charge, as compared to the zero 
B-field case. Hence in our setup, the model contains 48 
O3$^-$-planes and 16 O3$^+$-planes, and 3 O7$^-_3$-planes and one 
O7$^+_3$-plane. 

Gathering the two results together, the RR tadpole conditions read
\beqa
&\sum_a N_a n_a^1 n_a^2 {\tilde n}_a^3 + \frac 12 N_{\rm flux} -8= 0 
\nonumber\\
&\sum_a N_a n_a^1 m_a^2 m_a^3 +16 = 0 \nonumber\\
&\sum_a N_a m_a^1 n_a^2 m_a^3 +16 = 0 \nonumber\\
&\sum_a N_a m_a^1 m_a^2 {\tilde n}_a^3 +8 = 0 
\eeqa

Finally, there might be a subtle effect on the  3-form fluxes we are 
allowed to turn on. As noticed in \cite{fp,kst} for the toroidal 
orientifold case, the quantization conditions of 3-form fluxes depend on 
the kind of O3-planes in the configuration. There are four kinds of 
O3-planes, classified \cite{witbar} by the $\IZ_2$ value 
$(r_1,r_2)$ of the NSNS and RR 2-forms on an ${\bf RP}_2$ surrounding the 
O3-plane in the transverse 6d space. For instance, the O3$^-$-, 
O3$^+$-planes discussed above correspond to O3-planes with 
$(r_1,r_2)=(0,0)$, $(1/2,0)$, resp. If the integral of the NSNS (resp. RR) 
3-form field strength along one of the basis 3-cycles is integer but odd, 
consistency requires the corresponding 3-cycle to pass through an odd 
number of O3-planes with $r_1=1/2$ (resp. $r_2=1/2$).

We would like to provide an alternative derivation of this rule. For each 
basic 3-cycle $C$ in $\IT^6$, there is a 3-cycle ${\tilde C}$ closed on 
${\IT}^6/(\Omega R)$ which is not closed on $\IT^6$ and has half its 
volume. For instance, the 3-cycles given by ${\rm Re} z_1\in[0,1)$, ${\rm 
Re} z_2\in[0,1)$, ${\rm Re} z_3\in[0,1/2)$. If e.g. $\int_C H_3=$ odd, 
then $\int_{\tilde C}H_3=1/2$ mod $\IZ$. We would like to correlate this 
with the properties of the O3-planes through which ${\tilde C}$ and $C$ 
pass. To do this, remove a small 3-ball in ${\tilde C}$ around each 
O3-plane, whose boundary is an ${\bf RP}_2$, due to the orientifold 
quotient. We obtain a submanifold ${\tilde C}'$ with boundary, 
$\partial{\tilde C}'$ homologically given by  the sum of the classes of 
the ${\bf RP}_2$'s, denoted $\Sigma_i$. The $H_3$-flux $\int_{{\tilde C}'} 
H_3$ is however still equal to $\int_{\tilde C}H_3=1/2$ mod $\IZ$. From 
these conditions we have
\beqa
\int_{{\tilde C}'} H_3 = \int_{\partial{\tilde C}'} B_2 = 
\sum_i
\int_{\Sigma_i} B_i=\sum_i (r_1)_i 
\eeqa
Hence $\sum_i (r_1)_i=1/2 \, {\rm mod}\, \IZ$, which implies that the 
3-cycle passes through an odd number of O3-planes with $r_1=1/2$.

\medskip

One may expect that in our configurations, the presence of O3$^+$-planes
leads to modified quantization conditions. However, it is easy to verify 
that in the simplest case of non-zero B-field in just one two-torus, the 
configuration of O3-planes is such that any 3-plane passes through an even 
number of O3$^+$-planes. Therefore the correct quantization conditions 
require even $H_3$ and $F_3$ flux integrals over the basic 3-cycles
\footnote{One may worry about a possible correlation between such quanta 
and the RR charge of the O7-planes. However, the above geometric argument 
to show this correlation does not work, since the 3-cycles necessarily 
are not transverse to the O7-plane, and the boundaries of the small 
3-balls do not correspond to ${\bf RP}^2$'s located at the O7-planes.
Hence O7-planes seemingly do not modify the quantization conditions
imposed by the O3-plane distribution.}. Configurations with B-fields in 
several complex planes may however introduce modified quantization
conditions.

Let us provide one example of model with non-trivial B-field, just for 
illustration. Consider the D-brane configuration in table 
\ref{chiralthree}.

\begin{table}[htb] \footnotesize
\renewcommand{\arraystretch}{1.25}
\begin{center}
\begin{tabular}{|c||c|c|c|}
\hline
$N_a$ & $(m_a^1,n_a^1)$ & $(m_a^2,n_a^2)$ & $(m_a^3,{\tilde n}_a^3)$ \\
\hline\hline
$8$ & $(-1,1)$ & $(1,0)$ & $(-2,-1)$ \\
$4$ & $(-1,0)$ & $(1,-1)$ & $(2,1)$ \\
$2$ & $(0,1)$ & $(0,1)$ & $(0,1)$ \\
$12$ & $(1,0)$ & $(-1,0)$ & $(0,1)$\\
$12$ & $(1,0)$ & $(0,1)$ & $(-2,0)$ \\
\hline
\end{tabular}
\end{center}
\caption{\small D9-brane configuration for the model with B-flux}
\label{chiralthree}
\end{table}   

And let us saturate the RR tadpole with a flux $N_{\rm flux}=192$ and 
180 ${\ov D3}-branes$ like the above. Supersymmetry in the remaining 
branes is preserved for suitable choices of the  compactification areas. 
The chiral spectrum is
\beqa
& U(4)\times U(2) \times USp(2) \times USp(12) \times USp(12) & 
\nonumber\\
& 4(4,2;1,1,1)_{1,1}+ 2({\ov 4},1;2,1,1)_{-1,0} +2({\ov 
4},1;1,1,12)_{-1,0} + 4(6,1;1,1,1)_{-2,0}+ &\nonumber \\
& +4(10,1;1,1,1)_{2,0} + 2(1,2;2,1,1)_{0,1} + 4(1,1;1,1,1)_{0,2}+ 
4(1,3;1,1,1)_{0,-2} &\nonumber \\
& 180(4,1;1,1,1)_{1,0}+180(1,2;1,1,1)_{0,1}& 
\eeqa
Anomalies cancel using the additional contribution from the WZ terms. The 
model has a Pati-Salam gauge group (which can be 
subsequently broken to the Standard Model with correct hypercharge, by 
separating branes). However, the structure of families is not 
satisfactory, due to the contribution of the flux-induced WZ terms.
In any event, it provides a simple example of model building in this 
setup.

\section{T-dual version and D-branes on non-Calabi-Yau manifolds}
\label{noncy}

Models with NSNS and RR fluxes transform under T-duality in an interesting 
way. Denoting by $x$ the T-duality direction, the rough rules are as 
follows: i) RR forms transform by 
increasing or decreasing their degree for components transverse to or 
along $x$, respectively; ii) components of the NSNS 3-form field $H_3$ 
transverse to $x$ remain invariant; iii) finally, components of $H_3$ 
along $x$ (and two additional directions $y$, $z$) transform into 
components of metric curvature in the T-dual geometry, which forms a 
non-trivial bundle of the T-dual circle over the subspace 
parametrized by the coordinates $y$, $z$. The first Chern class of this 
bundle is given by the total flux of the original $H_3$ along $x$, $y$, 
$z$. 

The effect of T-duality has been studied quite generally in 
\cite{halflat}, and in the particular case of the torus in \cite{kstt}. It 
has been argued that geometries related by T-duality to Calabi-Yau 
compactifications with fluxes are neither Kahler nor complex, and hence 
not Calabi-Yau. In supersymmetric cases they however admit a 
(non-Levi-Civita) connection of $SU(3)$ holonomy. As discussed in 
\cite{halflat}, the geometry can be described (in a particular limit) as
a deformation of a Calabi-Yau space by a metric deformation characterized 
by some forms, which we refer to as `metric fluxes'. The study of general 
compactifications with metric fluxes (as well as other fluxes)
seems a promising direction towards understanding more general string 
compactifications \footnote{See \cite{halflat,torsion,kstt} for 
other recent work on this kind of compactification, and e.g. 
\cite{others} for applications to heterotic compactification.}.

\medskip

From this viewpoint, our models in this paper, upon application of 
T-duality, can be turned into the first (almost) supersymmetric chiral 
compactifications on non-Calabi-Yau manifolds with D-branes. These T-dual 
models are simpler to describe as deformations by metric fluxes of the 
T-dual toroidal orbifolds. Specifically, perform a T-duality along the 
horizontal direction $x^8$ in the third two-torus, denoted $x$ for 
shorthand. The dual model 
corresponds to type IIA compactified on a (metric flux deformed, see 
later) $\IT^6/(\IZ_2\times \IZ_2)$ orbifold, modded out by $\Omega R'$, 
with $R':x^a\to -x^a$, $a=4,5,6,7,9$, $R:x\to x$, where the
T-dual coordinate is also denoted $x$. Needless to say, our discussion 
applies even more 
simply to T-duality for toroidal orientifolds with fluxes, although in 
this case the complete configurations are non-supersymmetric.

The T-dual background includes RR 2- and 4-form field strength 
background, and NSNS 3-form background. In addition, 
components $(H_3)_{abx}$, introduce a non-trivial metric component
\beqa
g_{bx}\propto x^a
\eeqa
in the T-dual space, in fact turning the corresponding 3-torus (spanned by 
$x^a$, $x^b$, $x$) into a non-trivial $\IS^1$ bundle (with fiber 
parametrized by $x$) over the 2-torus spanned by $x^a$, $x^b$. 
Although the space is topologically different, locally on the base the 
geometry differs from a toroidal one by the 1-form
\beqa
g_{(x)}=\frac{g_{ax}}{g_{xx}} dx^a
\eeqa
which encodes the twisting \cite{kstt}. Its curvature 
$\omega_{(x)}=-dg_{(x)}$ is the curvature of the non-trivial bundle. 

The above orientifold quotient introduces a number of O4-planes, spanning 
$M_4$ and the direction $x$. Part of the corresponding RR tadpole is 
saturated by the fluxes. The remaining is saturated by D-branes, which in 
our supersymmetric case are D8-branes wrapped on the first two two-tori, 
times a 1-cycles in the third one. Regarding the T-dual background as a 
toroidal orbifold deformed by a metric flux, which is the right viewpoint 
from the results in \cite{halflat}, allows to discuss the D-brane 
configurations quite explicitly, almost as simply as in the absence of 
field strength fluxes.

\medskip

In the following we would like to discuss two important topological 
couplings which have not been mentioned in detail in the literature, 
although they play a crucial role in the construction of 
compactifications with metric fluxes, and the introduction of D-branes in 
them. In particular, we describe that combinations of  metric fluxes and 
RR fluxes contribute to RR tadpoles, and to Wess-Zumino terms in D-brane 
worldvolumes. Although motivated by our specific orbifold models, our 
analysis in this section is completely general and valid for other 
compactifications.

\subsection{RR tadpoles and metric fluxes}

T-duality implies that certain combinations of RR and metric fluxes 
are sources for certain RR potentials. Hence Type IIA compactified on 
a non-Calabi-Yau manifold $\IX_6$ corresponding to Calabi-Yau space
$\IY_6$ deformed with metric  fluxes encoded in a 1-form $g_{(x)}$ must 
contain Chern-Simons couplings responsible for this phenomenon. In this 
section we describe how they arise.

From the T-dualization of the type IIB coupling (\ref{cs}) along a 
component of $H_3$, the couplings we need are
\beqa
\int_{10d} \, g_{(x)} \wedge F_{4(x)} \wedge F_{6}\, =
\, \int_{10d} \,\omega_{(x)}\wedge F_{4(x)} \wedge C_5
\eeqa
where we have introduced the notation 
$(F_p)_{(x)}=F^{(p)}_{a_1\ldots a_{p-1}x}dx^{a_1} \ldots dx^{a_{p-1}}$. 
This coupling shows that RR 4-form and metric fluxes generate a tadpole 
for the RR 5-form potential.

We now show that this coupling simply arises from the expasion of the 
kinetic term for the RR 3-form on $\IX_6$, with metric $g$, around the
Calabi-Yau metric $g_0$, to first order in the metric flux 
deformation $\delta g$. We have the structure
\beqa
g_0=\pmatrix{g_{xx} & 0 \cr 0 & g_{ab}} \quad ; \quad
\delta g=\pmatrix{0 & g_{ax} \cr g_{ax} & 0}
\eeqa
and $g=g_0+\delta g$. Expanding a piece of kinetic term of the RR 
5-form on $\IX_6$ around $g_0$ to first order in $\delta g$, we have
\beqa
S_{\rm{kin.}g} & = & \int_{M_4\times \IX_6}\, F_{6(x)}\wedge *_{g} 
F_{6(x)} \, = \,
\int_{M_4\times \IX_6}\, \sqrt{g} \,\, F_{0123xa} \, F^{0123xa} \,
d({\rm Vol}) \, = \nonumber \\
&= & \int_{M_4\times \IX_6} \, \sqrt{g}\,\, F_{0123xa}\, 
F_{0123\mu\nu} \, g^{\mu x}\, g^{\nu a}\, d({\rm Vol})
\eeqa
where $\mu,\nu=4,\ldots, 9$. The zeroth order in $\delta g$ leads to the 
kinetic term of the 5-form in the background metric $g_0$ of $\IY_6$. The 
first order in $\delta g$ is obtained by taking $\sqrt{g}\to \sqrt{g_0}$, 
$\mu,\nu\neq x$. We obtain
\beqa
\Delta S & = & \int_{M_4\times \IY_6}\, \sqrt{g_0} \,\, F_{0123xa} 
F_{0123bc} \, g^{b x}\, g^{c a}\, d({\rm Vol})
\label{tocs}
\eeqa
We now use the Hodge duality relation in $\IY_6$, $F_6=*_{g_0} F_4$. 
Replacing
\beqa
F_{0123bc}\, =\, \sqrt{g_0}\, \epsilon_{0123bca_1 a_2 a_3 x}\,
(g_0)^{a_1b_1}\, (g_0)^{a_2b_2}\, (g_0)^{a_3b_3}\, (g_0)^{xx}\, 
F_{b_1 b_2 b_3 x}
\eeqa
into (\ref{tocs}), and using $g^{bx}=-(g_0)^{bd} g_{dx} (g_0)^{xx}$, we 
get
\beqa
\Delta S & = & -\int_{M_4\times \IY_6}\, (\det g_0) \, F_{0123xa}\,
F_{b_1b_2b_3x}\, (g_{(x)})_d \, \epsilon^{0123dab_1b_2b_3x} \,  
d({\rm Vol}) \nonumber \\
& =& \int_{M_4\times \IX_6} \, g_{(x)}\, \wedge F_{4(x)} \wedge 
F_{6}\, 
\eeqa
This generalizes to similar couplings for other RR forms.
The general conclusion is that the kinetic term for a RR form in a 
geometry deformed by metric fluxes leads to the kinetic term in the
undeformed metric plus a Chern-Simons coupling of the metric flux to RR 
fields.

\subsection{World-volume WZ terms and metric fluxes}

A similar analysis can be carried out for other important couplings in the 
presence of metric fluxes. For instance, the existence of world-volume WZ 
couplings for D9-branes in the presence of $H_3$, $F_3$ fluxes 
\cite{urangaflux}
\beqa
\int_{D9} B_2\wedge F_3 \wedge \left( \tr F^3 \right)^{(0)}=
\int_{D9} B_2 \wedge C_2 \wedge \tr F^3  
\eeqa
implies that D8-branes in the presence of RR 4-form and metric fluxes 
develop a WZ term
\beqa
\int_{D8} g_{(x)}\wedge F_{4(x)} \wedge \left( \tr F^3 \right)^{(0)}=  
\int_{D8} g_{(x)}\wedge C_{3(x)} \wedge \tr F^3   
\eeqa
Such terms indeed arise on D8-branes in the presence of metric fluxes. In
fact, they come from the Chern-Simons terms of D8-branes in $\IX_6$, 
\beqa
\int_{D8} P(C_3) \wedge \tr F^3
\eeqa
expanded around the configuration in $\IY_6$, taking into account that the 
pullback of $C_3$ has an expansion
\beqa
P(C_3)_{abc}=(C_3)_{abc}+\partial_c x\, (C_3)_{abx}
\eeqa
Here $x$ denotes the embedding of the D8-brane in the non-trivial $\IS^1$ 
bundle, and hence $\partial_a x=(g_{(x)})_a$. Thus the first term in the 
expansion gives
\beqa
\int_{D8} g_{(x)}\wedge C_{3(x)} \wedge \tr F^3 
\eeqa
explaining the required coupling.

Using these arguments it is straightforward to translate any model 
constructed in terms of $H_3$, $F_3$ fluxes into a model with non-trivial 
metric fluxes, at least for T-duality along a single direction. We expect 
much research in the construction of D-brane configurations in these 
non-Calabi-Yau spaces.

\section{Final comments}
\label{final}

Compactifications with field strength fluxes are a most promising avenue 
for model building, in that they provide a canonical mechanism to 
stabilize most moduli of the compactification. 

In this paper we have studied the construction of compactifications with 
D-branes, leading to chiral gauge sectors, and moduli stabilization by 
NSNS and RR fluxes. We have described different approaches to achieve this 
aim, and provided explicit examples with D3-branes at singularities (with 
supersymmetry broken in the closed string sector) and D-branes with 
world-volume magnetic fluxes (with a supersymmetric visible sector and a 
supersymmetry breaking hidden one). Along the 
way we have covered some new interesting properties of the models, like 
how to address the subtle behaviour of D-branes in the presence of NSNS 
fluxes, the relation between tadpoles and anomalies, etc.

Although we mainly centered in (almost) supersymmetric model building, our 
techniques can be inmediatly applied to the construction of 
non-supersymmetric models, for instance in toroidal orientifolds. We 
expect that within this less restrictive setup one can achieve the 
construction of far more phenomenologically appealing gauge sectors.

Many further directions remain open. In the context of moduli 
stabilization, it would be interesting to explore 
the interplay between the flux induced scalar potential with other 
sources of potential for moduli, like non-supersymmetric sets of D-branes, 
or non-perturbative corrections. This step is crucial in order to 
understand the fate of the moduli which are not stabilized by the fluxes.
It would also be interesting to determine patterns for the values at which 
moduli stabilize, in order to understand for instance what properties the 
underlying model must have in order to lead to e.g. small 4d gauge 
couplings, or large radii. 

In the context of generalizing these constructions, a nice extension of 
our models would be the construction of compactifications with fluxes and 
chiral D-branes in more general Calabi-Yau spaces. Also, as we have 
discussed, compactification with fluxes are closely related to 
compactifications on non-Calabi-Yau manifolds, related to metric fluxes. 
Study of the latter is one of the most exciting directions in generalizing 
our knowledge of string theory vacua. Introduction of D-branes in them 
would largely enhance our model building possibilities, bringing us 
perhaps one step closer to understanding the structure of the observed 
physics.

\centerline{\bf Acknowledgements}

We thank M. Cvecti\u{c}, G. Honecker, F. Marchesano and F. Quevedo for 
useful discussions, 
and especially G. Aldazabal and L. Ib\'a\~nez for fruitul conversations at 
the first stages of this project. A.M.U. thanks M.~Gonz\'alez for kind 
encouragement and support. J.G.C wants to thank M. P\'erez for her 
patience and affection. This work has been partially supported by CICYT 
(Spain). The research of J.G.C. is supported by the Ministerio de 
Educaci\'on, Cultura y Deporte through a FPU grant. 

\newpage

\appendix

\section{NSNS and RR fluxes in a type IIB orientifold with D3-branes at 
singularities.}
\label{zthree}

In this appendix we would like to discuss the construction of orientifold 
of type IIB on $\IT^6/\IZ_3$ by the orientifold action $\Omega R$ with 
$R:z_i\to -z_i$, and NSNS and RR 3-form field strength fluxes. 
This is a T-dual version of the model in \cite{abpss}. As 
discussed in the main text, the models turn out be non-supersymmetric in 
the closed string sector since the orbifold and the $\IZ_3$ invariant 
fluxes necessarily preserve different supersymmetries.

A direct way of constructing would be to introduce fluxes in the
$\IT^6/\IZ_3$ orientifold, which will necessarily be non-supersymmetric,
but should at least be imaginary self-dual with respect to the metric of
the 6d internal space.

We prefer instead to take an indirect route, which takes more advantage of
our discussions in the main text, and
carry out the construction as follows. We start with type IIB on 
$\IT^6/\IZ_3$ modded out by $\Omega R$, and introduce a 3-form flux of the 
form
\beqa
G_3\, =\, 2\, dz_1dz_2d{\ov z}_3
\label{z3flux}
\eeqa
This flux stabilizes moduli at a factorized product of three two-tori 
(i.e. off-diagonal Kahler parameters are frozen to zero) with 
complex structure parameters $\tau_i=e^{2\pi i/3}$, and stabilizes the 
dilaton at $\phi=e^{2\pi i/3}$. It is also properly quantized over 3-cycles
in the torus, with integrated fluxes giving even numbers.
Notice that this flux preserves some of the supersymmetries of the
underlying ${\IT}^6/(\Omega R)$ geometry. Concretely it preserves the
spinor $\xi_0$ (satisfying $\gamma^{\ov i}\xi_0=0$ in the above complex
coordinates), as well as the spinors $\gamma^1\gamma^3\xi_0$ and $\gamma^2
\gamma^3\xi_0$ (hence preserves 4d $\NN=3$ supersymmetry). Being
supersymmetric the flux is automatically imaginary
self-dual with respect to the underlying metric.
For the above flux we have $N_{\rm flux}=12$, hence cancellation of RR
tadpoles requires the introduction of 20 D3-branes.

We now mod out the configuration by the $\IZ_3$ orbifold generated 
by 
\beqa
\theta: (z_1,z_2,z_3) \to (e^{2\pi i/3} z_1, e^{2\pi i/3} z_2, e^{4\pi 
i/3} z_3)
\eeqa
namely $z_i\to e^{2\pi v_i}z_i$ with $v=1/3(1,1,2)$.

For this quotient to
be possible it is crucial that the flux (\ref{z3flux}) is invariant under
the action of $\theta$, so it corresponds to a possible flux in $\IT^6/\IZ_3$
(fulfills condition i) in section \ref{orbiflux}).
It is also important to notice that the $\IZ_3$ quotient of $\IT^6/(\Omega R)$
does not contain closed 3-cycles which are not closed in $\IT^6/\IZ_3$.
hence proper quantization is not spoilt
(point ii in section \ref{orbiflux}). Also, the collapsed cycles at
$\IZ_3$ singularities are 2- and 4-cycles, hence do not impose additional
quantization constraints (point iii in \ref{orbiflux}). Finally, the flux
is imaginary self-dual in the metric of the orbifold space (at the orbifold
point in moduli space), since the metric is inherited from that of $\IT^6$
(point iv in \ref{orbiflux}). Finally, the spinor preserved by the
orbifold projection is $\gamma^3\xi^0$, which is not any of the
supersymmetries preserved by the flux. Hence the final model is not
supersymmetric.

In this model, supersymmetry broken in the closed string sector by
the interplay between the orbifold projection and the flux. It is
possible that this kind of breaking of supersymmetry has some particularly
nice features, since the interactions between untwisted modes is
sensitive to supersymmetry breaking only via effects involving twisted
modes. It would be interesting to analyze the impact of this property on
the violations of the no-scale structure of the low energy supergravity
effective theory for these models (i.e. the degree of protection
against $\alpha'$ or $g_s$ corrections).

\medskip

To define the model completely, we need to specify the configuration of
the 20 D3-branes, which are still required to cancel the untwisted RR
tadpole. 
In the $\Omega R$ orientifold of $\IT^6$ there is one point, the origin
$(0,0,0)$, fixed under $\Omega R$ and $\theta$. At this point, cancellation
of RR twisted tadpoles requires the presence of D3-branes, with a
Chan-Paton matrix satisfying
\beqa
\Tr \, \gamma_{\theta,3}=-4
\eeqa
In addition, there are other $26$ points fixed under $\theta$ (and gathered
in 13 pairs under $\Omega R$), where there is no twisted RR tadpole. If
D3-branes are present, they should have traceless Chan-Paton matrix. Finally
there are 63 points fixed under $\Omega R$ (gathered in 21 trios under
$\IZ_3$) at which we may locate any number (even or odd) of D3-branes.

A simple solution would be to locate the 20 D3-branes at the origin, with
\beqa
\gamma_{\theta,3}=\diag (\id_4,e^{2\pi i/3} \id_8, e^{4\pi i/3} \id_8)
\eeqa
leading to an $\NN=1$ supersymmetric sector (to leading approximation,
since interactions with the closed sector would transmit supersymmetry
breaking), with spectrum
\beqa
& \NN=1 \; {\rm vect.mult.}\quad \quad & SO(4)\times U(8) \nonumber \\
& \NN=1 \; {\rm ch. mult.}\quad & 3\, [ \, 
(\fund,\fund)+(1,\ov{\Yasymm})\, ]
\eeqa

A more interesting possibility, which we adapt from \cite{lpt}, is to
locate 11 D3-branes at the origin, with
\beqa
\gamma_{\theta,3}=\diag (\id_1,e^{2\pi i/3} \id_5, e^{4\pi i/3} \id_5)
\eeqa
This leads to a gauge sector with
\beqa
& \NN=1 \; {\rm vect.mult.}\quad \quad & U(5) \nonumber \\
& \NN=1 \; {\rm ch. mult.}\quad & 3\, (\,5+{\ov{10}}\,)
\eeqa
One should now be careful in locating the additional D3-branes.
Introducing an odd number of D3-branes on top of the O3-plane at the
origin implies that it is an $\widetilde{O3}^-$-plane in notation of
\cite{witbar}, i.e. there exists a $\IZ_2$ $B_{RR}$ background on an
${\bf RP}_2$ around the O3-plane.
In order to be consistent with the fact that our flux has
even integral over the different 3-cycles implies that there should exist
other $\widetilde{O3}^{-}$-planes in the configuration.
The conditions in \cite{fp,kst}
state that for any 3-plane over which the integrated flux of $H_3$ is even
(any 3-plane in our case), the number of $\widetilde{O3}^{-}$-planes must 
be
even. The configuration in \cite{lpt} turns out to satisfy the corresponding
consistency conditions (the underlying reason being that it is consistent
for zero fluxes, which is a particular case of even quanta on 3-cycles).

To adapt this configuration, denote $A$, $B$ or $C$ the coordinate of an
O3-plane in a complex plane, according to whether $z_i=1/2$,
$z_i=e^{2\pi i/3}/2$ or $z_i=(1+e^{2\pi i/3})/2$. The remaining 9 D3-branes
in the model are located on top of O3-planes at the points
\beqa
& (A,A,A)\quad (B,B,C) \quad (C,C,B) \nonumber \\
& (A,0,0) \quad (B,0,0)\quad (C,0,0) \nonumber \\
& (0,A,0) \quad (0,B,0) \quad (0,C,0)
\eeqa
This set is invariant under exchange of fixed points by $\IZ_3$, and
introduces the right number of $\widetilde{O3}^{-}$-planes at the right 
places.
The additional D3-branes do not lead to additional gauge symmetries.

Thus the final model contains a 3-family $SU(5)$ GUT gauge sector (although
without adjoint chiral multiplets to break it down to the Standard Model),
as the only gauge sector of the theory. In addition, its closed string
sector is non-supersymmetric. It would be interesting to estimate the
impact of the supersymmetry breaking on the gauge sector \cite{grana}.
It would also be interesting to construct other models based on the
$\IZ_3$ orbifold, or other orbifold models. We leave these interesting
question for future work.

\section{K-theory discussion}
\label{ktheory}

\subsection{Wrapped D-branes with non-zero closed string field strength 
fluxes}

Configurations of D-branes in the presence of NSNS field strength 
fluxes are subtle. Stable D-brane configurations are characterized by
their charges, which are classified by a suitable K-theory group. In 
simple situations, the K-theory classes are in one-to-one correspondence 
with cohomological classes, and stable D-branes can be characterized by 
specifying the homology cycle they wrap (and the Chern classes of the 
gauge bundle they carry). In other situations, K-theory may differ from 
cohomology and it may not be possible to wrap a D-brane on a non-trivial 
homology cycle, or a D-brane in a non-trivial homology cycle may decay and 
disappear. In the presence of NSNS 3-form field strength it has been 
proposed that the K-theory group classifying D-brane charges is the 
twisted K-theory group $K_{[H]}$ of spacetime \cite{wittenkth}. This has 
been argued mainly when the cohomology class $[H_3]$ is torsion 
\cite{fw,kapustin}, and there have been attempts at defining it properly 
for non-torsion $[H_3]$ \cite{wittenktwo}. This has proved quite 
difficult, even conceptually (e.g. definition of a D-brane system as the 
final state of annihilation of spacetime filling brane-antibrane pairs 
seems to require an infinite number of such parent pairs).

On the other hand, an alternative operational definition of the set of 
allowed D-brane charges in a configuration, including the presence of NSNS 
field strength was provided in \cite{mms}, based on a physical 
understanding of instanton processes mediating D-brane decays (see
\cite{evslin} for a recent discussion). The two basic rules 
\footnote{We are
considering a situation with trivial Stiefel-Witney classes, pertinent to
our applications.} to classify topological D-brane charges on a spacetime
$X_9\times \IR$ with NSNS 3-form field strength $H_3$ are

i) The configuration given by a D-brane wrapping a homologically
non-trivial cycle $W$ is consistent, unless the pullback of the bulk
3-form field strength $H_3$ onto $W$ defines a cohomologically non-trivial
class in $W$.

ii) A configuration of a D$p$-brane wrapped on a homologically non-trivial
$p$-cycle $W$ (and propagating in time)
is unstable to decay into the vacuum if there exists a
$(p+3)$-cycle $W'$, containing $W$, such that the pullback of $H_3$ onto
$W'$ is the Poincare dual of the class of $W$ in $W'$.

\begin{figure}
\begin{center}
\centering
\epsfysize=3.5cm
\leavevmode
\epsfbox{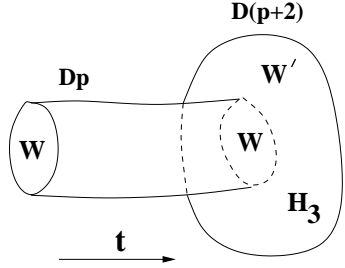}
\end{center}
\caption[]{\small 
A D$p$-brane ending on a D$(p+2)$-brane with volume along some component 
of $H_3$. This represents a consistent way of wrapping a D$(p+2)$-brane on 
a cycles with non-zero $H_3$ flux, or a process mediating the decay of a 
D$p$-brane wrapped on a homologically non-trivial cycle $W$.}
\label{dinst}
\end{figure}

The decay of the D-branes in ii) is mediated by the following instanton
process: Due to i) we cannot wrap a D-brane in $W'$. If we wrap an
euclidean D$(p+2)$-brane on it, the inconsistency can be cured by
introducing a D$p$-brane wrapped on $W$ (and propagating in time) and
ending on it. The D$(p+2)$-brane on $W'$ therefore is an instanton
mediating the decay of the D$p$-brane on $W$. The process is shown in
figure \ref{dinst}.

The above rules can be explained in more pedestrian terms with explicit
examples. To derive i) from physical considerations, take a D3-brane
wrapped in a 3-cycle $\Sigma$ such that $\int_\Sigma H_3=k$. On the
D3-brane worldvolume there is a coupling
\beqa
\int_{\Sigma_3\times \IR} H_3\wedge {\tilde A}_1
\eeqa
where ${\tilde A}_1$ is the gauge potential dual to the world-volume
1-form gauge field. This implies that the $H_3$ background generates a
world-volume tadpole for ${\tilde A}_1$, which renders the theory
inconsistent (since it is not possible to satisfy the equations of motion
for ${\tilde A}_1$. The inconsistency is avoided if there are $k$
D1-branes which span the direction $\IR$ above, and other direction, in
which they are of semi-infinite extent because they end on the D3-brane.
The dimensions are described by

\begin{center}
\begin{tabular}{ccccccccccc}
D3 & 0 & $\times$ & $\times$ & $\times$ & $\times$ & $\times$ & $\times$
& 7 & 8 & 9 \\
D1 & 0 & $\times$ & $\times$ & $\times$ & $\times$ & $\times$ & $\dashv$
& $\times$ & $\times$ & $\times$ \\
\end{tabular}
\end{center}

where $\dashv$ denotes that the brane is  semi-infinite in the 
corresponding 
direction.
The boundary of the D1-branes on the D3-brane contributes $k$ units of
charge under ${\tilde A}_1$, cancelling the world-volume tadpole.
Regarding the direction 6 as time, this configuration described  above
represents the decay of $k$ D1-branes via a D3-brane instanton.
The configuration is easily T-dualized into a set of D$p$-branes spanning
$\IR^p$ times a semi-infinite line, ending on a D$(p+2)$-brane spanning
$\IR^p\times \Sigma$. The boundary of the D$p$-branes cancel the
contribution from $H_3$ to the world-volume dual gauge potential
${\tilde A}_p$.

These physically motivated rules, as well as their dual versions, provide
an operational definition of conserved D-brane charges which we exploit
in the analysis of our configurations.

\subsection{D9-branes wrapped on 3-cycles with non-zero $H_3$}
\label{extrabranes}

Our configurations in the present paper contain D9-branes wrapped on
the internal space, on which we have turned on non-trivial field strength
fluxes. Since the pullback of $H_3$ into the D9-brane volume is non-zero,
the configurations are strictly speaking not consistent as they stand.
We need to introduce additional D-branes `ending' on the D9-branes to
render the latter consistent.
Notice that the fact that our configurations do not contain net D9-brane
charge (in a sense contain D9- and anti-D9-branes\footnote{In a
supersymmetry preserving fashion due to their worldvolume magnetic
fields.}) does not avoid the inconsistency, which appear for each D9-brane
independently.

\begin{figure}
\begin{center}
\centering
\epsfysize=3.5cm
\leavevmode
\epsfbox{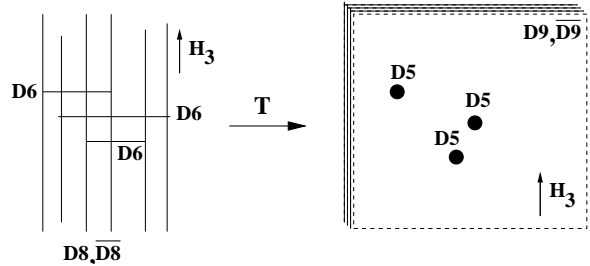}
\end{center}
\caption[]{\small A system D8 - (anti-D8)-branes with volume along some 
component of $H_3$ is consistent if there are additional D6-branes 
starting and ending on them. The T-dual version is that a system of D9 -
(anti-D9)-branes with volume along some component of $H_3$ is consistent 
if additional fractional D5-branes are present.}
\label{dghost}
\end{figure}

To understand what additional D-branes we need, consider the simplified
situation of D9- anti-D9-brane pairs feeling just one component of $H_3$,
say along 789. T-dualizing along, say 6, this configuration is equivalent
to D8 - anti-D8-branes along 012345789 with $H_3$ along 789. The latter
configuration can be made consistent by adding sets of $k$ D6-branes
starting from each anti-D8-brane and ending on each D8-brane, as shown
in figure \ref{dghost}. T-dualizing back to D9-branes, the required 
additional
D-branes are D5-branes, sitting at a point in the directions 789 where
$H_3$ is non-trivial, and in an additional direction. The fact that the
T-dual D6-branes had finite extent and stretched between D8-branes implies
the D5-branes must be `fractional' in the sense of \cite{merons}.

Hence in our discussion in this paper we assume implicitly that
additional branes of this kind have been added to our configurations,
to render them consistent
\footnote{The role of the additional branes is even richer. D-branes
in the presence of $H_3$ flux carry in principle not standard gauge
bundles, but twisted gauge bundles \cite{wittenkth}, defined in terms of
a cocyle related to the class of $H_3$. This is analogous to saying that
$H_3$ is magnetically charged under the world-volume gauge potential, so
the latter is not globally well defined. Because of the additional
branes, with magnetic charge cancelling the above, the twisting can be
safely ignored (equivalently, gauge fixed to have support at the location
of the additional brane), and one can work with usual world-volume bundles.
This is essential in order to define magnetised D-branes in the familiar
fashion.}. Since our flux configurations are more
involved, we do not specify the additional brane content in detail.

However, we need not be too specific about it. We would like to 
emphasize that their presence does not alter any of the main properties of 
our interest. First, given that the D9-brane content and the fluxes 
preserve supersymmetry, we expect the additional branes to do so as well. 
A more precise argument would be as follows: Consider, in the absence of 
closed string fluxes, a $D9-{\ov{D9}}$ pair, supersymmetric due to their 
world-volume gauge bundles. Open strings stretched between them lead to 
4d $\NN=1$ chiral multiplets, with scalars parametrizing a flat direction 
along which the $D9-{\ov{D9}}$-brane form a(n irreducible) bound state, 
preserving the same supersymmetry as before. Consider now re-introducing 
the closed string fluxes, preserving that supersymmetry. The 
D-brane configuration is unchanged, i.e. no additional D-branes are 
required, since the D-brane bound state carries zero D9-brane charge, and 
hence does not wrap a 3-cycles with $H_3$ flux. The D-brane configuration 
is therefore supersymmetric along the complete flat direction, and so must 
be at the origin, where the bound state becomes reducible into the 
configuration of a D9-brane, a ${\ov{D9}}$-brane, and the additional 
branes.

Second, it is easy to
realize that the additional branes do not introduce additional chiral
matter in the spectrum of the theory. This is manifest in the picture of
D6-branes ending on D8-branes, where the intersection point is known to
preserve 8 supercharges, and lead only to vector-like matter. Therefore
the main properties of the models can be obtained ignoring the subtlety
of the presence of the additional branes, as we do in the main text
\footnote{The presence of additional branes was unnoticed in the
discussion of other configurations of D-branes and fluxes in the
literature \cite{urangaflux,strongcp}. Happily their addition does not 
modify the chiral
content and anomaly cancellation mechanisms there discussed.}

\subsection{Instantons violating homological D-brane charge}
\label{instanton}

Another striking feature of D-branes in the presence of $H_{3}$ NSNS 
3-form field strength is that there exist processes violating conservation 
of the homological charge carried by the D-branes, via ii).

This phenomenon occurs for D-brane charge associated to classes point-like
in the directions spanned by $H_3$.
In our models, the only net homology charges carried by the D-branes 
are D3-brane charge and D7-brane charge. Only the D3-brane charge is 
pointlike in the directions spanned by $H_3$, hence there are processes 
changing the D3-brane charge. Indeed, for each independent 3-cycle 
$\Sigma$ over which $\int_\Sigma H_3=k$, there exist an `instanton', given 
by an euclidean D5-brane wrapped on $\Sigma$ and spanning the three 
non-compact spatial dimensions (and at some fixed time $t_0$), which 
changes the number of D3-branes by $k$ units.

Since the amount of D3-brane charge was fixed by cancellation of the RR 
tapdoles generated by the O3-planes, one may wonder how the latter is 
cancelled after the transition. The answer follows form noticing that the 
D5-brane instanton couples magnetically to $B_{RR}$, hence
\beqa
dF_3=\delta(D5)
\eeqa
where $\delta(D5)$ is a bump 4-form localized on the volume of the 
D5-brane. We then see that the amount of $F_3$ flux along the cycle 
$\tilde{\Sigma}$ dual to $\Sigma$ changes by $k$ units in the D3-number 
violating process. 
\beqa
\Delta\int_{\tilde{\Sigma}} F_3=
\int_{t=t_0-\delta t} \Delta\int_{\tilde{\Sigma}} F_3
-\int_{t=t_0+\delta t} \Delta\int_{\tilde{\Sigma}} F_3
=\int_{\Pi_4} dF_3=\int_{\Pi_4} \delta(D5)=1 
\eeqa
where $\Pi_4$ is $\Sigma$ times the interval $[t_0-\delta t,t_0+\delta 
t]$. This implies that the contribution to the $C_4$ tadpole from 
$N_{\rm flux}$ changes by $k$ units
\beqa
\Delta\int_{X_6} H_3 F_3=\int_{X_6} H_3 \Delta F_3=\int_{\Sigma} H_3 
\int_{\tilde{\Sigma}} F_3 = k
\eeqa
therefore compensates the disappearance of $k$ D3-branes. 

Some remarks are in order

$\bullet$ The above description suggests that NS5-brane instantons can 
mediate the decay of D3-branes in the presence of RR 3-form field 
strength flux $F_3$. This is an additional modification not taken into 
account in the usual K-theory groups discussed in the literature, due to 
the NS nature of the instantonic brane. Certainly the classification of
D-brane states in the presence of general fluxes is still an open question 
and deserves further study.

$\bullet$ RR tadpole cancellation works as usual, taking into account the 
charges 
carried by the fluxes. This also implies that instanton processes mediated 
by euclidean D-branes in the presence of $H_{NS}$ do violate D-brane 
number, but conserve the charges under the RR fields.

$\bullet$ We conclude by pointing out that the above instantons can be 
regarded as domain walls in time. In fact, some of them are supersymmetry 
preserving, and should connect different supersymmetric configurations of 
fluxes and D-branes. In fact, using one of the above spatial coordinates 
as time, our above instanton processes become spatial domain walls, 
separating regions of space with different fluxes and D-branes, of 
the kind studied in \cite{klst}. 

$\bullet$ In conclusion the above instantons in our models simply 
represent possible tunneling processes connection our supersymmetric 
configurations with other supersymmetric configurations. Hence the 
instanton processes, and more generally the possibility of D-brane 
disappearance due to the fluxes, can be safely ignored in our context.
(It would be an open issue in non-supersymmetric models, where an analysis 
of the energetics of the barrier would be requrired to ensure enough 
metastability).


\newpage

\begin{thebibliography}{99}

\bibitem{vw}
C. Vafa, E. Witten, `On orbifolds with discrete torsion' 
J. Geom. Phys. 15 (1995) 189, hep-th/9409188; see also
A. Font, L. E. Ibanez, F. Quevedo, `$Z_N\times Z_M$ orbifolds and discrete 
torsion', Phys. Lett. B217 (1989) 272,1989. 

\bibitem{fluxes}
J.~Polchinski, A.~Strominger, `New vacua for type II string theory',
Phys. Lett. B388 (1996) 736, hep-th/9510227; \\
K.~Becker, M.~Becker, `M theory on eight manifolds', Nucl. Phys. B477
(1996) 155, hep-th/9605053; \\
J.~Michelson, `Compactifications of type IIB strings to four-dimensions
with nontrivial classical potential', Nucl. Phys. B495
(1997) 127, hep-th/9610151; \\
S.~Gukov, C.~Vafa, E.~Witten, `CFT's from Calabi-Yau four folds',
Nucl. Phys. B584 (2000) 69, Erratum-ibid. B608 (2001) 477, hep-th/9906070; 
\\
K.~Dasgupta, G.~Rajesh, S.~Sethi, `M theory, orientifolds and G-flux',
JHEP 9908 (1999) 023, hep-th/9908088; \\
S. Gukov, `Solitons, superpotentials and calibrations',
Nucl. Phys. B574 (2000) 169, hep-th/9911011;\\
T.~R.~Taylor, C.~Vafa, `R R flux on Calabi-Yau and partial supersymmetry
breaking', Phys. Lett. B474 (2000) 130, hep-th/9912152; \\
K. Behrndt, S. Gukov, `Domain walls and superpotentials from M theory on 
Calabi-Yau three folds', Nucl. Phys. B580 (2000) 225, hep-th/0001082;\\
B.~R.~Greene, K.~Schalm, G.~Shiu, `Warped compactifications in M and F
theory', Nucl. Phys. B584 (2000) 480, hep-th/0004103; \\
G.~Curio, A.~Klemm, D.~Lust, S.~Theisen, `On the vacuum structure of type
II string compactifications on Calabi-Yau spaces with H fluxes',
Nucl. Phys. B609 (2001) 3, hep-th/0012213; \\
M.~Haack, J.~Louis, `M theory compactified on Calabi-Yau fourfolds with 
background flux', Phys. Lett. B507 (2001) 296, hep-th/0103068; \\
J.~Louis, A.~Micu, `Type 2 theories compactified on Calabi-Yau threefolds 
in the presence of background fluxes', hep-th/0202168; \\

 


\bibitem{gkp}
S.~B.~Giddings, S.~Kachru, J.~Polchinski, `Hierarchies from fluxes in
string compactifications', hep-th/0105097.

\bibitem{fp}
A.~R.~Frey, J.~Polchinski, `N=3 warped compactifications',
hep-th/0201029.

\bibitem{kst}
S.~Kachru, M.~Schulz, S.~Trivedi, `Moduli stabilization from fluxes in a
simple iib orientifold', hep-th/0201028.

\bibitem{tt}
P. K. Tripathy, S. P. Trivedi, `Compactification with flux on K3 and 
tori', hep-th/0301139.

\bibitem{ferrara}
R. D'Auria, Sergio Ferrara, S. Vaula, `N=4 gauged supergravity and a IIB 
orientifold with fluxes', New J.Phys. 4 (2002) 71, hep-th/0206241;\\
S. Ferrara, M. Porrati, `N=1 no-scale supergravity from IIB orientifolds',
Phys. Lett. B545 (2002) 411, hep-th/0207135; 

\bibitem{witbar}
E. Witten, `Baryons and branes in anti-de Sitter space',
JHEP 9807 (1998) 006, hep-th/9805112.

\bibitem{bbhl}
K. Becker, M. Becker, M. Haack, J. Louis, 
`Supersymmetry breaking and alpha-prime corrections to flux induced 
potentials', JHEP 0206 (2002) 060, hep-th/0204254. 

\bibitem{kklt}
S. Kachru, R. Kallosh, A. Linde, S. P. Trivedi, `De sitter vacua in string 
theory', hep-th/0301240.

\bibitem{gp}
M. Gra\~na, J. Polchinski, `Supersymmetric three form flux perturbations 
on AdS$_5$', Phys. Rev. D63 (2001) 026001, hep-th/0009211. 

\bibitem{dm}
M.~R.~Douglas, G.~W.~Moore, `D-branes, quivers, and ALE instantons',
hep-th/9603167; \\
M.~R.~Douglas, B.~R.~Greene, D.~R.~Morrison, `Orbifold resolution by
D-branes', Nucl. Phys. B506 (1997) 84, hep-th/9704151.  

\bibitem{singu}
G.~Aldazabal, L.~E.~Ib\'a\~nez, F.~Quevedo, A.~M.~Uranga,
`D-branes at singularities: A Bottom up approach to the string embedding
of the standard model', JHEP 0008 (2000) 002, hep-th/0005067; \\
D.~Berenstein, V.~Jejjala, R.~G. Leigh, `The Standard model on a D-brane',
Phys. Rev. Lett. 88 (2002) 071602, hep-ph/0105042;\\
L.~F.~Alday, G.~Aldazabal, `In quest of just the standard model on
D-branes at a singularity', JHEP 0205 (2002) 022, hep-th/0203129.

\bibitem{bgkl}
R.~Blumenhagen, L.~Goerlich, B.~Kors, D.~Lust, `Noncommutative
compactifications of type I strings on tori with magnetic background
flux', JHEP 0010 (2000) 006, hep-th/0007024.
 
\bibitem{afiru}
G.~Aldazabal, S.~Franco, L.~E.~Ibanez, R.~Rabadan, A.~M.~Uranga, `D=4
chiral string compactifications from intersecting branes',
J. Math. Phys. 42 (2001) 3103, hep-th/0011073; `Intersecting brane
worlds', JHEP 0102 (2001) 047, hep-ph/0011132.

\bibitem{bkl}
R.~Blumenhagen, B.~Kors, D.~Lust, `Type I strings with F flux and B flux',
JHEP 0102 (2001) 030, hep-th/0012156; \\

\bibitem{rest}
L.~E.~Ibanez, F.~Marchesano, R.~Rabadan, `Getting just the standard model
at intersecting branes', JHEP 0111 (2001) 002, hep-th/0105155; \\
S.~Forste, G.~Honecker, R.~Schreyer, `Orientifolds with branes at angles',
JHEP 0106 (2001) 004, hep-th/0105208; \\
R.~Blumenhagen, B.~Kors, D.~Lust, T.~Ott, `The standard model from stable
intersecting brane world orbifolds', Nucl. Phys. B616 (2001) 3,
hep-th/0107138; \\
D.~Cremades, L.~E.~Ibanez, F.~Marchesano, `SUSY Quivers, Intersecting
Branes and the Modest Hierarchy Problem', hep-th/0201205; `Intersecting 
brane models of particle physics and the Higgs mechanism' hep-th/0203160;
`Standard Model at Intersecting D5-branes: lowering the 
string scale', hep-th/0205074. \\
C.~Kokorelis, `GUT model hierarchies from intersecting branes',
hep-th/0203187; `New standard model vacua from intersecting branes',
hep-th/0205147.

\bibitem{more}
R.~Blumenhagen, V.~Braun, B.~K\"ors, D.~L\"ust, `Orientifolds of K3 and
Calabi-Yau manifolds with intersecting D-branes', JHEP 0207 (2002) 026,
hep-th/0206038; \\
A. M. Uranga, `Local models for intersecting brane worlds', 
hep-th/0208014.

\bibitem{susy}
M.~Cvetic, G.~Shiu, A.~M.~Uranga, `Chiral four-dimensional N=1
supersymmetric type 2A orientifolds from intersecting D6 branes',
Nucl. Phys. B615 (2001) 3, hep-th/0107166; `Three family supersymmetric
standard - like models from intersecting brane worlds', Phys. Rev. Lett.
87 (2001) 201801, hep-th/0107143.

\bibitem{susy2}
R.~Blumenhagen, L.~Gorlich, T.~Ott, `Supersymmetric intersecting branes on 
the type 2A T6 / Z(4) orientifold', hep-th/0211059;\\
M.~Cvetic, I.~Papadimitriou, G.~Shiu, `Supersymmetric three family SU(5) 
grand unified models from type IIA orientifolds with intersecting 
D6-branes', hep-th/0212177;\\
G. Honecker, `Chiral supersymmetric models on an orientifold of $Z_4 
\times Z_2$ with intersecting D6-branes', hep-th/0303015.

\bibitem{orbif}
R.~Blumenhagen, L.~Gorlich, B.~Kors, `Supersymmetric 4-D orientifolds of
type IIA with D6-branes at angles', JHEP 0001 (2000) 040, hep-th/9912204; \\
S.~Forste, G.~Honecker, R.~Schreyer, `Supersymmetric Z(N) x Z(M)
orientifolds in 4-D with D branes at angles', Nucl. Phys. B593 (2001)
127, hep-th/0008250.
 
\bibitem{reviews}
A. M. Uranga, `Chiral four-dimensional string compactifications with 
intersecting D-branes', hep-th/0301032; \\
R.~Blumenhagen, V.~Braun, B.~Kors, D.~Lust, ` The standard model on the 
quintic', hep-th/0210083.

\bibitem{halflat}
S. Gurrieri, J. Louis, A. Micu, D. Waldram, `Mirror symmetry in 
generalized Calabi-Yau compactifications', hep-th/0211102.

\bibitem{torsion}
G.L. Cardoso, G. Curio, G. Dall'Agata, D. Lust, G. Zoupanos, `NonKahler 
string backgrounds and their five torsion classes', hep-th/0211118;
S. Gurrieri, A. Micu, `Type IIB theory on half-flat manifolds',
hep-th/0212278 

\bibitem{kstt}
S. Kachru, M. B. Schulz, P. K. Tripathy, S. P. Trivedi, 
`New supersymmetric string compactifications', hep-th/0211182.

\bibitem{others}
K. Becker, K. Dasgupta, `Heterotic strings with torsion', JHEP 
0211 (2002) 006, hep-th/0209077; K. Becker, M. Becker, K. Dasgupta, P. S. 
Green,`Compactifications of heterotic theory on nonKahler complex 
manifolds. 1', hep-th/0301161.

\bibitem{magnetised}
C.~Bachas, `A Way to break supersymmetry', hep-th/9503030;
C.~Angelantonj, I.~Antoniadis, E.~Dudas, A.~Sagnotti, `Type I strings on
magnetized orbifolds and brane transmutation', Phys. Lett. B489 (2000)
223, hep-th/0007090; G. Pradisi, `Magnetized (shift)orientifolds', 
hep-th/0210088.

\bibitem{rabadan}
R. Rabadan, `Branes at angles, torons, stability and supersymmetry',
Nucl. Phys. B620 (2002) 152, hep-th/0107036.

\bibitem{bdl}
M.~Berkooz, M.~R.~Douglas, R.~G.~Leigh, `Branes intersecting at angles',
Nucl. Phys. B480 (1996) 265, hep-th/9606139.     

\bibitem{wittenkth}
E. Witten, `D-branes and K theory', JHEP 9812 (1998) 019,
hep-th/9810188.

\bibitem{urangakth}
A. M. Uranga, `D-brane probes, RR tadpole cancellation and K theory 
charge', Nucl. Phys. B598 (2001) 225, hep-th/0011048.
  
\bibitem{fernando}
F. Marchesano, private communication.

\bibitem{urangaflux}
A.~M.~Uranga, `D-brane, fluxes and chirality', JHEP 0204 (2002) 016,
hep-th/0201221.

\bibitem{strongcp}
G. Aldazabal, L. E. Ibanez, A. M. Uranga, `Gauging away the strong CP 
problem', hep-ph/0205250.

\bibitem{imr}
See first reference in \cite{rest}; \\
I. Antoniadis, E. Kiritsis, J. Rizos, `Anomalous U(1)s in type 1 
superstring vacua', Nucl. Phys. B637 (2002) 92, hep-th/0204153. 

\bibitem{yuk}
D. Cremades, L.E. Ibanez, F. Marchesano, `Towards a theory of quark 
masses, mixings and CP violation', hep-ph/0212064; `Yukawa couplings in 
intersecting d-brane models', hep-th/0302105.

\bibitem{witvec}
E. Witten, `Toroidal compactification without vector structure',
JHEP 9802 (1998) 006, hep-th/9712028.

\bibitem{abpss}
C. Angelantonj, M. Bianchi, G. Pradisi, A. Sagnotti, Ya.S. Stanev, `Chiral 
asymmetry in four-dimensional open string vacua',
Phys. Lett. B385 (1996) 96, hep-th/9606169.

\bibitem{lpt}
J. Lykken, E. Poppitz, S. P. Trivedi, `Branes with GUTs and supersymmetry 
breaking', Nucl. Phys. B543 (1999) 105, hep-th/9806080.

\bibitem{grana}
M. Grana, `MSSM parameters from supergravity backgrounds',
hep-th/0209200.

\bibitem{fw}
D. S. Freed, E. Witten, `Anomalies in string theory with D-branes',
hep-th/9907189.

\bibitem{kapustin}
A. Kapustin, `D-branes in a topologically nontrivial B 
field', Adv. Theor. Math. Phys. 4 (2000) 127, hep-th/9909089.

\bibitem{wittenktwo}
E.Witten, `Overview of K theory applied to strings',
Int. J. Mod. Phys. A16 (2001) 693, hep-th/0007175.

\bibitem{mms}
J.M. Maldacena, G. W. Moore, N. Seiberg, `Geometrical interpretation of 
D-branes in gauged WZW models', JHEP 0107 (2001) 046, hep-th/0105038.

\bibitem{evslin}
J. Evslin, `Twisted k-theory from monodromies', hep-th/0302081.

\bibitem{merons}
J. Brodie, `Fractional Branes, Confinement, and Dynamically Generated 
Superpotentials', Nucl. Phys. B532 (1998) 137, hep-th/9803140.

\bibitem{klst}
S. Kachru, X. Liu, M. B. Schulz, S. P. Trivedi, `Supersymmetry changing 
bubbles in string theory', hep-th/0205108.

\bibitem{recent}
R. Blumenhagen, D. Lust, T. R. Taylor, `Moduli Stabilization in Chiral 
Type IIB Orientifold Models with Fluxes', hep-th/0303016.

\end{thebibliography}
\end{document}